\newcommand*{\FINAL}{}

\documentclass[sigconf,nonacm]{acmart}

\usepackage{algorithmic}
\usepackage{graphicx}
\usepackage{textcomp}
\usepackage{xcolor}
\usepackage{subcaption}
\usepackage{booktabs}
\usepackage{natbib}
\usepackage{multirow}
\usepackage{balance}
\usepackage{soul,listings,lstautogobble}
\usepackage{color,colortbl}
\usepackage[frozencache=true,cachedir=minted-cache]{minted}
\usepackage{paralist}
\usepackage{booktabs}
\usepackage{caption}
\usepackage{tabularx}
\usepackage[normalem]{ulem} 
\usepackage{array}
\usepackage{capt-of}   
\usepackage[most]{tcolorbox}

\tcbset{
    colback=gray!10,
    colframe=gray!60!black,
    boxrule=0.4pt,
    arc=3pt,
    left=6pt,right=6pt,top=6pt,bottom=6pt
}

\newcolumntype{Y}{>{\raggedright\arraybackslash}X}

\definecolor{ListBGColor}{rgb}{0.95,0.95,0.95}
\definecolor{KeywordColor}{rgb}{0,0,0.6}
\definecolor{ListCommentColor}{rgb}{0.4, 0.27, 0.173}
\makeatletter
\newcommand\labelline[1]{%
    \def\@currentlabel{\thelstnumber}\label{#1}}
\renewcommand\lineref[1]{
    \@ifundefined{r@#1}{0}{\ref{#1}}
}
\makeatother
\makeatletter
\providecommand*{\NAT@spacechar}{~}
\makeatother

\newcommand{\remove}[1]{}

\ifdefined \REVISION
    
    \renewcommand{\remove}[1]{{\color{purple} \sout{{#1}}}}
    
\else
\fi

\newcommand{\todo}[1]{}
\newcommand{\mawad}[1]{}
\newcommand{\mosama}[1]{}
\newcommand{\brandon}[1]{}
\newcommand{\alex}[1]{}
\newcommand{\ryan}[1]{}
\newcommand{\future}[1]{}
\newcommand{\note}[1]{}

\ifdefined \FINAL
\else
\renewcommand{\todo}[1]{\par\noindent{\color{red}\textbf{TODO:} #1}\par}
\renewcommand{\mawad}[1]{\par\noindent{\color{blue}\textbf{mawad:} #1}\par}
\renewcommand{\mosama}[1]{\noindent{\color{violet}\textbf{mosama:} #1}}
\renewcommand{\brandon}[1]{\par\noindent{\color{cyan}\textbf{Brandon:} #1}\par}
\renewcommand{\alex}[1]{\par\noindent{\color{lime}\textbf{Alex:} #1}\par}
\renewcommand{\ryan}[1]{\par\noindent{\color{teal}\textbf{Ryan:} #1}\par}
\renewcommand{\future}[1]{\par\noindent{\color{brown}\textbf{Future:} #1}\par}
\renewcommand{\note}[1]{\par\noindent{\color{brown}\textbf{Note:} #1}\par}

\fi

\AtBeginDocument{%
  \providecommand\BibTeX{{%
    \normalfont B\kern-0.5em{\scshape i\kern-0.25em b}\kern-0.8em\TeX}}}

\begin{document}

\title{{Iris}: First-Class Multi-GPU Programming Experience in Triton}

\author{Muhammad Awad}
\email{muhaawad@amd.com}
\orcid{0000-0002-6914-493X}
\affiliation{%
       \institution{Advanced Micro Devices, Inc.}
       \city{Santa Clara, CA}
       \country{USA}
}

\author{Muhammad Osama}
\email{muhammad.osama@amd.com}
\orcid{0000-0003-1616-6817}
\affiliation{%
       \institution{Advanced Micro Devices, Inc.}
       \city{Santa Clara, CA}
       \country{USA}
}

\author{Brandon Potter}
\email{brandon.potter@amd.com}
\orcid{0009-0001-0223-1641}
\affiliation{%
       \institution{Advanced Micro Devices, Inc.}
       \city{Austin, TX}
       \country{USA}
}

\renewcommand{\shortauthors}{Awad, Osama and Potter}

\hyphenation{Bulk-sync-hronous}
\hyphenation{point-er-based}

\begin{abstract}
    Multi-GPU programming traditionally requires developers to navigate complex trade-offs between performance and programmability. High-performance implementations typically rely on low-level HIP/CUDA communication libraries that demand substantial engineering effort for even basic overlap patterns, while simpler abstractions often sacrifice performance. We present Iris, a multi-GPU communication library implemented entirely in Python and Triton that eliminates this trade-off. Iris provides tile-based symmetric memory abstractions that naturally align with Triton's programming model, enabling developers to write single-source kernels that seamlessly interleave computation and communication. We demonstrate a taxonomy of compute-communication overlap patterns—from bulk-synchronous to fine-grained workgroup specialization—that can be implemented with minimal code changes in Iris, often requiring just a few additional lines within the same Triton kernel. Our evaluation shows that Iris achieves near-optimal bandwidth utilization in microbenchmarks and delivers up to 1.79$\times$ speedup over PyTorch and RCCL for GEMM+All-Scatter workloads, demonstrating that high-level implementations can match or exceed heavily-optimized libraries while dramatically simplifying multi-GPU programming.
\end{abstract}


\maketitle
\section{Introduction}
\label{sec:intro}

Modern AI workloads demand near-peak performance to extract the full efficiency of AI systems. Teams of specialists with deep understanding of both model characteristics and hardware architecture are required to craft highly optimized training and inference kernels for these AI workloads. Even for seasoned engineers, this process requires iterative refinement, hardware-specific tuning, and extensive experimentation. This challenge arises because contemporary AI models comprise numerous operators, each with multiple potential optimization strategies. Determining the appropriate optimizations depends on a wide range of factors—including model structure, configuration parameters, hardware vendor and generation, system topology, and supporting software ecosystems. Moreover, end-to-end performance is strongly influenced by distributed-parallel execution primitives—such as all-reduce and all-to-all—that orchestrate computation across devices.

Distributed parallelism further amplifies the complexity. Practitioners routinely combine data, tensor, pipeline, and expert parallelism strategies and expect these hybrid approaches to perform consistently across heterogeneous hardware platforms. For system and library developers, this presents a significant challenge: kernel efficiency is tightly coupled to network characteristics and system architecture, both of which vary substantially in real-world deployments. As a result, collective communication libraries must deliver high performance across diverse and evolving environments.

Given this complex landscape, higher-level abstractions are essential to simplify development without sacrificing performance. However, modern accelerators also introduce specialized units and mechanisms—such as tensor cores and asynchronous memory-movement engines (SDMA, TMA, and asynchronous copy instructions)—that demand fine-grained, tile-based programming to fully exploit their capabilities for overlapping communication and computation. A number of efforts have advanced the state of the art, including compiler-driven approaches (e.g., XLA~\cite{OpenXLA:2025:XLA}, TVM~\cite{Chen:2018:TAE}, and Triton~\cite{Tillet:2019:TAI}) as well as template-based libraries (e.g., CUTLASS, CuTe~\cite{Kerr:2017:CUTLASS}, ThunderKittens~\cite{Spector:2024:TSA}). Among these, Triton has shown sustained maturity, performance portability, and broad adoption for computation. However, a critical gap remains: while these abstractions have successfully tackled local operators and compute kernels, they have largely overlooked communication as an equally important concern. As distributed training and inference scales to hundreds or thousands of GPUs, communication becomes the dominant performance bottleneck.

\begin{figure}[t]
    \centering
    \begin{subfigure}[b]{0.48\columnwidth}
        \centering
        \includegraphics[trim=125 175 175 155, clip, width=\textwidth]{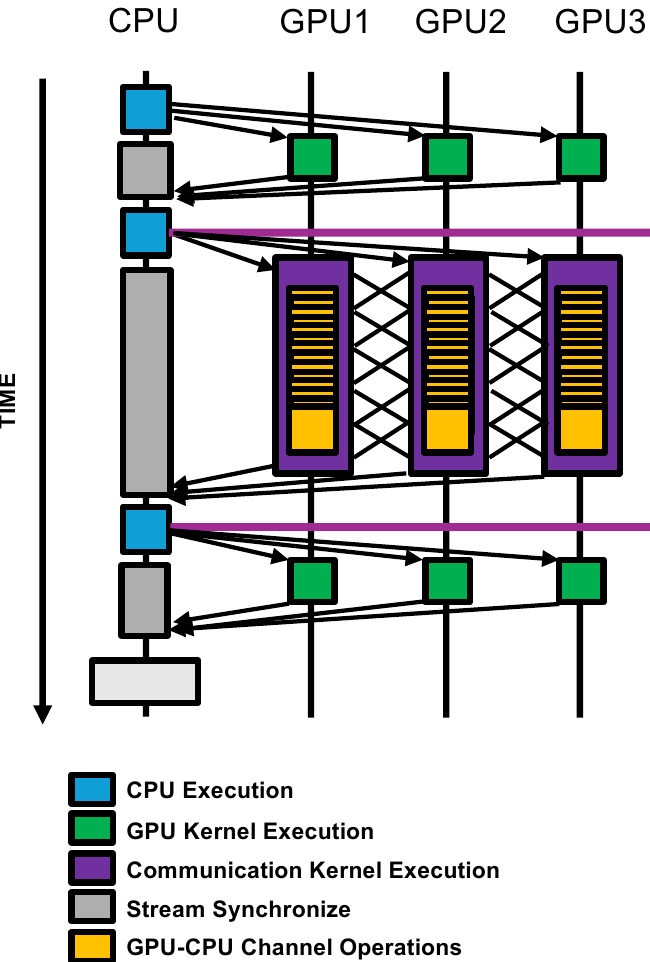}
        \caption{Bulk-Synchronous}
        \label{fig:bsp}
    \end{subfigure}
    \hfill
    \begin{subfigure}[b]{0.48\columnwidth}
        \centering
        \includegraphics[trim=125 175 175 155, clip, width=\textwidth]{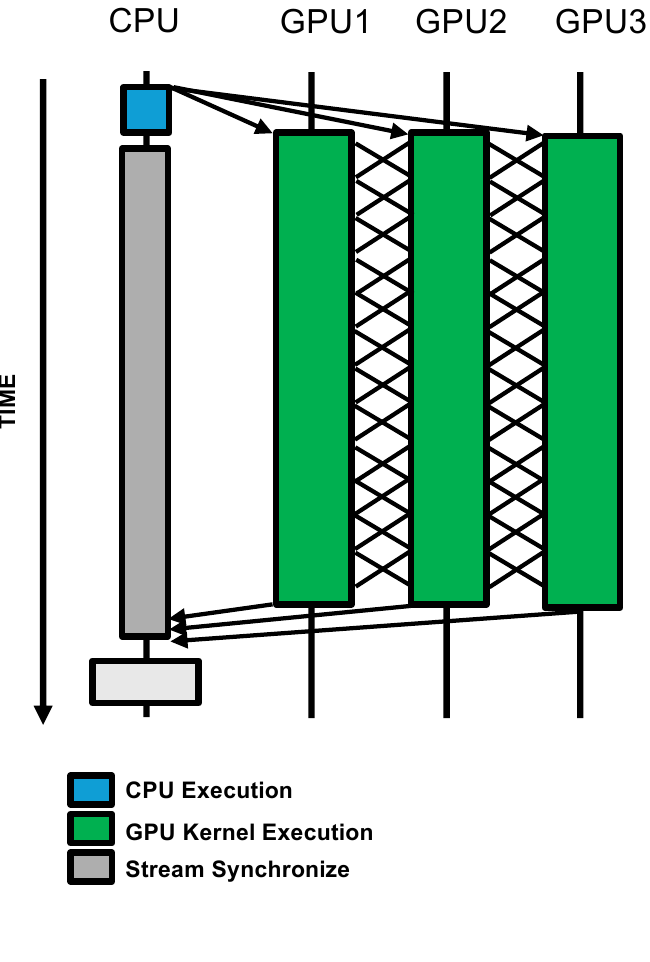}
        \caption{Fine-Grained Overlap}
        \label{fig:fg}
    \end{subfigure}
    \caption{Execution model comparison: (a) bulk-synchronous execution with rigid sequential phases and CPU-initiated kernel launches versus (b) fine-grained overlap enabling dynamic communication at tile granularity with GPU-initiated execution.}
    \label{fig:bsp_vs_finegrained}
\end{figure}

\paragraph{The Consensus: Fine-Grained Overlap is Essential.} The research and production communities have converged on a clear solution: overlap communication with computation at fine granularity. Rather than executing communication and computation in rigid, sequential bulk-synchronous phases (Figure~\ref{fig:bsp}), modern workloads demand fine-grained overlap where data is communicated as soon as it is produced, at tile granularity (Figure~\ref{fig:fg}), so that computation can proceed on other tiles without waiting. This fine-grained overlap hides communication latency behind useful work, eliminating the idle ``bubbles'' that plague bulk-synchronous execution. Recent systems such as TorchTitan's AsyncTP~\cite{Liang:2025:TOS} and production LLM training pipelines~\cite{DeepSeekAI:2025:DVT,Spector:2025:LMN,Trifan:2025:EMT} have demonstrated the necessity of this approach.

\paragraph{Communication: The Missing Abstraction.} Fine-grained overlap requires a communication abstraction that operates at the same tile granularity as Triton's computational model. However, Triton provides no such abstraction. While Triton's tile-based programming model has revolutionized how developers write optimized compute kernels—automatically handling memory coalescing, shared memory management, and intra-kernel scheduling--communication remains an afterthought. Developers must rely on external libraries (RCCL~\cite{AMD:2025:RCL}, NCCL~\cite{NVIDIA:2025:NCL}) that operate at coarse kernel boundaries, or hand-craft device-to-device data movement without compiler support. This approach forces communication to remain outside the compiler's purview, preventing the tile-level interleaving that the consensus approach demands. The result is that fine-grained overlap, despite being widely recognized as essential, remains inaccessible to Triton developers.

\paragraph{The Challenge: Architectural Limitations of Existing Approaches.} Recognizing this gap, several efforts have attempted to bring communication into Triton. However, these attempts face fundamental architectural constraints. Current efforts typically wrap vendor libraries (rocSHMEM~\cite{AMD:2025:ROS}, NVSHMEM~\cite{NVIDIA:2025:NVSHMEM}) as opaque bytecode, linking them into Triton kernels. While pragmatic, this approach introduces several limitations. First, these libraries inherit design constraints from the OpenSHMEM specification and carry technical debt from APIs originally designed for CPU-based distributed computing, which do not align naturally with Triton's tile-centric programming model and introduce anti-patterns that work against compiler optimizations. Second, and more critically, wrapping external libraries as opaque binaries prevents the compiler from seeing communication operations, precluding co-optimization of computation and communication, intelligent scheduling, and unified fusing of kernels across their boundaries. Communication remains a second-class citizen—linked as binary blobs rather than first-class Triton code. Similarly, traditional collective communication libraries (CCLs) such as RCCL and NCCL impose bulk-synchronous semantics with CPU-initiated kernel launches, adding coordination overhead, kernel teardown costs, and redundant memory transfers at kernel boundaries. These architectural constraints limit the ability to achieve the native, compiler-visible, tile-granular primitives required for true fine-grained overlap.

\paragraph{Iris: Native Communication for Tile-Based Programming.} We present Iris~\cite{Awad:2025:IFM}\footnote{Iris is open-source and available at \url{https://github.com/ROCm/iris}}, the first multi-GPU library architected from the ground up for Triton's tile-based programming model to full enable the fine-grained computation and communication overlap within AI workloads (Figure~\ref{fig:fg}). Unlike existing approaches that wrap legacy SHMEM libraries as opaque bytecode, Iris is implemented entirely in Python and Triton, giving the compiler full visibility into both computation and communication. Iris provides native symmetric memory abstractions that enable developers to write concise single-source kernels that seamlessly interleave computation and communication at tile granularity with no external dependencies, no opaque library calls, no bulk-synchronous phases. Iris also offers an experimental Gluon backend using Triton's \texttt{@gluon.jit} and \texttt{@aggregate} decorators for improved ergonomics, but this paper focuses on the standard Triton API for clarity and broader compatibility. Our contributions are as follows:
\begin{itemize}
    \item \textbf{Native Triton implementation:} The first multi-GPU communication library implemented entirely in Python and Triton, providing full compiler visibility and enabling co-optimization of computation and communication
    \item \textbf{Tile-based symmetric memory API:} Pythonic abstractions that align naturally with Triton's tile-centric programming model, supporting both value-based and pointer-based communication primitives
    \item \textbf{Taxonomy of fused patterns:} A comprehensive classification of compute-communication overlap strategies, including bulk-synchronous, producer-consumer, and workgroup-specialized approaches
    \item \textbf{Performance validation:} Experimental evaluation demonstrating up to 1.79$\times$ speedup over PyTorch and RCCL for GEMM+All-Scatter workloads across multiple problem sizes
    \item \textbf{Open-source release:} Fully open-source implementation enabling reproducibility and community adoption
\end{itemize}
\section{Background and Related Work}
\label{sec:background}

Iris builds on established GPU communication mechanisms—symmetric memory, direct inter-GPU interconnects, and well-defined memory consistency models—while introducing a novel programming abstraction that distinguishes it from prior work. We first review the underlying hardware and runtime infrastructure, then survey related efforts to integrate communication into GPU programming frameworks.

\subsection{Background}
We briefly review the key mechanisms that enable multi-GPU communication on AMD platforms. This includes the physical interconnect topology, the runtime interfaces for establishing symmetric memory across processes, and the memory consistency model that provides correctness guarantees. Together, these components form the execution substrate on which Iris is built.

\subsubsection{Interconnect Topology}
Iris targets scale-up environments where multiple GPUs within a single node are connected via a high-bandwidth interconnect. The AMD Instinct~MI300X and~MI325X platforms~\cite{AMD:2025:ACA} use seven high-bandwidth, low-latency AMD Infinity Fabric links per GPU to form a fully connected 8-GPU system. Each GPU is also connected to the host CPU via a x16 PCIe Gen 5 link. This fully connected mesh topology provides direct peer-to-peer access between any GPU pair without traversing the host CPU, enabling the low-latency, high-bandwidth communication that Iris leverages for efficient multi-GPU operations. Iris exploits this topology to implement efficient collective operations and point-to-point communication patterns directly in Triton kernels.

\subsubsection{Symmetric Memory via IPC}
\label{sec:ipc}

Iris establishes symmetric memory across GPUs using HIP's inter-process communication (IPC) mechanism~\cite{AMD:2025:HIP}. Each process allocates device memory using standard \texttt{hipMalloc}, then exports handles via \texttt{hipIpcGetMemHandle} and imports peer handles via \texttt{hipIpcOpenMemHandle}. This enables direct memory access across GPU boundaries: each GPU can read from and write to any peer GPU's memory using simple pointer arithmetic. Iris uses coarse-grained memory semantics for simplicity and portability, relying on the memory consistency model (described below) to ensure correctness of cross-GPU operations. By leveraging IPC, Iris exposes a clean symmetric memory abstraction to programmers, enabling Triton kernels to perform remote memory operations as naturally as local ones.

\subsubsection{Memory Model and Synchronization}
\label{sec:rc}

Iris relies on AMD's memory model to provide formal correctness guarantees in multi-GPU execution. This model has been described in publicly available literature~\cite{HSA:2018:PSA} and is implemented concretely in AMDGPU LLVM~\cite{LLVM:2025:AUM}. The AMD memory model is Sequentially Consistent Heterogeneous Race Free (SC-HRF), analogous to the C++ model but extended with GPU-specific memory scopes. The model supports standard C++ memory orderings (\texttt{acquire}, \texttt{release}, \texttt{acq\_rel}, and \texttt{seq\_cst}) with familiar semantics: acquire operations prevent subsequent loads and stores from being reordered before them, while release operations prevent preceding loads and stores from being reordered after them. Critically, the model introduces hierarchical memory scopes—\texttt{wavefront} (warp), \texttt{workgroup} (block), \texttt{agent} (device), and \texttt{system}—that define visibility domains for synchronization operations. Triton already exposes these memory orderings and scopes through its atomic operations API, and Iris leverages Triton's implementation to use hardware-level synchronization primitives directly. For multi-GPU communication, Iris uses \texttt{agent}-scoped atomics to synchronize between GPUs within a node, and \texttt{system}-scoped operations when broader visibility is required.

Iris adopts this memory model because it is well-established, widely-adopted across CPU and GPU programming, and provides programmers with familiar, intuitive synchronization primitives (acquire/release, memory scopes) that are straightforward to reason about and use, while still delivering provable correctness guarantees for multi-GPU coordination.

\subsubsection{Symmetric Memory Programming Models}
\label{sec:shmem-background}

The symmetric memory programming model, established by the OpenSHMEM specification and implemented by hardware vendors as xSHMEM variants (rocSHMEM~\cite{AMD:2025:ROS} and NVSHMEM~\cite{NVIDIA:2025:NVSHMEM}), provides a foundational abstraction where each process allocates memory in a symmetric heap that is directly accessible by all peers. This model enables one-sided communication primitives—such as remote puts, gets, and atomic operations—that can be issued without requiring explicit receiver-side coordination, making it well-suited for GPU programming where fine-grained producer-consumer patterns are common.

Iris adopts this symmetric heap abstraction as its core memory model, recognizing its proven utility for multi-GPU communication. However, rather than wrapping existing xSHMEM implementations, Iris reimagines the programming interface with a modern, Pythonic API built natively in Triton. This approach preserves the conceptual benefits of symmetric memory while eliminating the legacy constraints of C-style interfaces and explicit thread-ID management that were inherited from CPU-era HPC programming models.

\subsection{Related Work}

\label{sec:related}

A number of systems provide abstractions for multi-GPU communication, symmetric memory, and fused compute-communication execution models. Table~\ref{tab:iris-comparison} contrasts Iris with existing approaches, highlighting two fundamental implementation strategies: native implementations architected for the target programming model versus wrapper-based approaches that integrate external libraries. Below, we discuss the most closely related efforts in detail.

\begin{table}[t]
       \centering
       \caption{Comparison of multi-GPU communication approaches.}
       \label{tab:iris-comparison}
       \small
       \resizebox{\columnwidth}{!}{%
              \begin{tabular}{lll}
                     \toprule
                     \textbf{Aspect}              & \textbf{Built for Triton} & \textbf{Wrapper-Based}              \\
                     \midrule
                     \textbf{Libraries}           & Iris                      & Triton-Dist., PyTorch               \\
                     \midrule
                     \textbf{Approach}            & Native Triton             & Wraps xSHMEM                        \\
                     \midrule
                     \textbf{Programming Model}   & Triton-native             & OpenSHMEM-based                     \\
                     \midrule
                     \textbf{Compiler Visibility} & Full, co-optimization     & Opaque, limited                     \\
                     \midrule
                     \textbf{API Style}           & Pythonic, tile-based      & C-style, Python-wrapped             \\
                     \midrule
                     \textbf{Language}            & Python + Triton           & Python + xSHMEM                     \\
                     \midrule
                     \textbf{Memory Model}        & C++/HIP model             & Ill-defined                         \\
                     \midrule
                     \textbf{Synchronization}     & Acquire/release semantics & \texttt{quiet}/\texttt{wait\_until} \\
                     \bottomrule
              \end{tabular}%
       }
\end{table}

\subsubsection{HIP/CUDA/C++-Based Approaches}

\paragraph{xSHMEM Libraries}
As discussed in Section~\ref{sec:shmem-background}, xSHMEM libraries (NVSHMEM, rocSHMEM) implement the OpenSHMEM specification~\cite{OpenSHMEM:2012:SPE} and establish the symmetric memory abstraction that Iris builds upon. However, their APIs do not align well with Triton's tile-centric programming paradigm: they require explicit thread-ID management and rely on low-level C-style interfaces. These abstractions were originally built for HPC-CPU environments and later ported to GPU, which limits their effectiveness for modern GPU programming models. Iris retains the symmetric heap abstraction but provides a modern, Pythonic API that integrates naturally with Triton's tile-based execution model.

\paragraph{Flux}
\citet{Chang:2024:FFS} introduced Flux, which targets symmetric memory communication but is implemented directly in CUDA and CUTLASS. While offering highly optimized kernels, this approach requires longer development cycles and relies heavily on C++ template metaprogramming. Iris maintains a simpler Python- and Triton-based design that enables faster prototyping, easier debugging, and full compiler visibility.

\subsubsection{Triton-Based Wrappers}

\paragraph{Triton-Distributed}
\citet{Zheng:2025:TPO} introduced Triton-Distributed, which wraps existing xSHMEM libraries behind Triton-specific Python APIs, introducing proof-of-concept fused kernels for GEMM and AllGather and MoE-style patterns. While similar in motivation to Iris, it inherits the limitations of its underlying SHMEM implementation: the communication layer remains a thin wrapper around vendor libraries, preventing compiler-level visibility. In contrast, Iris implements remote memory operations directly in Triton with no external dependencies, enabling full compiler visibility for fine-grained fused operations.

\paragraph{PyTorch Symmetric Memory and TorchTitan AsyncTP}
Recent PyTorch efforts~\cite{PyTorch:2025:2.9} introduce symmetric memory support at the framework level, enabling asynchronous tensor-parallel communication via decomposed point-to-point operations (\texttt{put}/\texttt{get}). TorchTitan leverages this for fused GEMM-AllGather and MoE patterns optimized via \texttt{torch.compile}. However, similar to Triton-Distributed, these systems rely on vendor-supplied communication backends, limiting compiler optimization opportunities. Iris draws inspiration from the decomposed pattern approach but exposes primitives directly within Triton kernels, enabling tile-level fused kernels and eliminating kernel-switching overhead while avoiding reliance on external communication libraries.

\section{Iris}
\label{sec:iris}

Iris is a multi-GPU library built from scratch for scaling with minimal dependencies (only Triton, PyTorch, and HIP runtime). The library provides intuitive and simple APIs for developers without requiring knowledge of distributed systems architecture, enabling Python and Triton developers to write multi-GPU code leveraging high-level language abstractions. First we will discuss the design decisions that influenced Iris's design and successfully resulted in an abstraction that allows Triton developers to be productive and use familiar abstractions.

\subsection{Design Philosophy}

Iris adopts the symmetric heap abstraction from SHMEM (as discussed in Section~\ref{sec:shmem-background}) but modernizes the programming model for GPU computing. Rather than porting legacy SHMEM APIs, we provide Pythonic and Triton-native interfaces that respect modern programming paradigms. As summarized in Table~\ref{tab:iris-comparison}, this distinguishes Iris from wrapper-based approaches that rely on existing xSHMEM implementations.

\subsubsection{Adoption of Symmetric Heap Abstraction}
Iris implements a Symmetric Heap abstraction. While Iris deviates from SHMEM-like library APIs, it implements one of the core ideas adopted in the OpenSHMEM specification: the symmetric heap. Symmetric heaps are simple to understand, implement, and use. The symmetric heap design provides predictable memory layouts across all GPUs, enabling efficient pointer translation with minimal overhead since each rank's memory has an identical structure at corresponding offsets.

\subsubsection{Familiar Memory Model and Pythonic APIs}
Rather than introducing new synchronization primitives or memory semantics, Iris adopts the well-established C++/HIP/CUDA memory model with acquire/release ordering (as detailed in Section~\ref{sec:rc}). This design choice is intentional: GPU programmers already reason about memory consistency, ordering, and synchronization in their single-GPU kernels. By reusing these familiar semantics for multi-GPU communication, Iris eliminates the need to learn a new memory model.

Both HIP/HSA~\cite{HSA:2018:PSA} and CUDA~\cite{NVIDIA:2025:CCP} programming models define clear semantics for atomic operations with configurable memory ordering (relaxed, acquire, release, acquire-release) and synchronization scopes (block, GPU, system). These well-defined scopes allow developers to precisely control the visibility and ordering of memory operations across different granularities---from thread block synchronization to system-wide coherence across GPUs. This familiarity extends to both host-side APIs (PyTorch-compatible tensor operations) and device-side APIs (Triton-native operations), which we detail in subsequent sections.

\subsubsection{Pure Python and Triton Implementation}
A key distinguishing feature of Iris is its implementation: the entire framework is built from scratch in Python and Triton without requiring external communication libraries or custom runtime dependencies. Unlike wrapper-based approaches such as Triton-Distributed (discussed in Section~\ref{sec:related}) that rely on rocSHMEM bytecode or other low-level communication primitives, Iris leverages only the existing PyTorch ecosystem (for host-side operations) and HIP runtime APIs (for GPU IPC, as described in Section~\ref{sec:ipc}). This design choice provides several advantages: (1) portability across different GPU vendors without vendor-specific communication libraries, (2) ease of debugging and modification since the entire codebase is in high-level Python and Triton, (3) simplified deployment with no additional system dependencies beyond PyTorch and the standard HIP/CUDA runtime, and (4) compiler visibility---the Triton compiler has full visibility into the entire codebase, enabling optimizations across computation and communication boundaries rather than treating communication primitives as opaque binary blobs linked into the final executable.

\subsubsection{Value- and Pointer-based APIs}
Existing SHMEM-based APIs treat the source and destination arguments to a SHMEM function as buffers that are pointed to by a pointer, along with their respective buffer sizes. CPU threads are more \textit{heavyweight} and typically work on buffers of data, which likely influenced this design choice. However, Iris targets GPUs where hundreds of thousands of threads are actively doing work. Massively-parallel GPUs require both value-based and pointer-based operations. Value-based data movement copies data directly from registers to other GPUs. In contrast, pointer-based data movement acts as a data copy between the main memory of local GPUs and that of another GPU.

The rise of tile-based programming frameworks such as Triton, ThunderKittens~\cite{Spector:2024:TSF}, and CuTe DSL~\cite{NVIDIA:2024:CTC} demonstrates the importance of value-based APIs that directly operate on tensors rather than raw bytes. These frameworks prioritize high-level tensor abstractions because they align with how developers reason about computation and data movement in modern GPU programming. Iris's value-based APIs enable developers to express operations at the granularity of computational tiles, moving partial results directly from registers to remote memory without intermediate buffering. This approach is particularly effective for fine-grained computation-communication overlap patterns where data becomes available incrementally during computation. We will provide examples that leverage both value-based and pointer-based APIs in Section~\ref{sec:patterns}.

\subsection{Host-Side APIs}

Iris provides Pythonic PyTorch-like host APIs organized into several categories: initialization, memory management, rank/world queries, host-side communication, and tensor construction. Iris implements a full symmetric heap, where each allocation returns a PyTorch tensor that wraps the allocated virtual memory address range. We organize the discussion into three main areas: core infrastructure setup, distributed operations, and tensor management. Table~\ref{lst:iris_host} at the end provides a quick reference summary of all host-side APIs.

\subsubsection{Core Infrastructure}

\paragraph{Constructor and Initialization.} The initialization process follows several key steps: (1) PyTorch Distributed initialization and rank assignment, (2) GPU device selection based on rank, (3) symmetric heap initialization on the selected device, (4) IPC handle creation and exchange across all ranks using PyTorch Distributed all-gather operations, (5) opening of remote IPC handles to establish cross-GPU memory access, and (6) creation of a tensor containing all heap base addresses for device-side translation. This setup enables seamless remote memory access through the translate function, which converts local pointers to remote addresses by computing offsets and applying them to destination heap bases. Iris sets the GPU device and treats each single GPU as its own rank in the distributed communicator\footnote{To simplify distributed programming, typically, each GPU is treated as its own rank rather than a single compute node.}.

\subsubsection{Distributed Operations}

\paragraph{Rank and Device Queries.} Iris provides several query functions for distributed computing context. The \texttt{get\_rank()} method returns the current process's rank ID in the distributed communicator, while \texttt{get\_num\_ranks()} returns the total number of ranks (world size). The \texttt{get\_heap\_bases()} method returns a tensor containing symmetric heap base addresses for all ranks, which is essential for device-side pointer translation.

\paragraph{Host-side Communication.} Iris provides two primary communication primitives. The \texttt{barrier()} function synchronizes all ranks across the entire system by first calling \texttt{torch.cuda.synchronize()} (or \texttt{stream.synchronize()} if a stream is specified) to ensure the local GPU has finished all queued work, then performing a global distributed barrier so all ranks reach the same point before proceeding. The \texttt{broadcast()} function broadcasts a value from one rank to all others, automatically detecting the value type and using the appropriate mechanism: for tensors and arrays it uses efficient PyTorch distributed tensor collectives, while for scalars and other objects it uses object broadcast. This intelligent type detection makes broadcasting seamless across different data types.

\subsubsection{Tensor Management}

\paragraph{Tensor Construction.} Since remote memory operations require symmetric heap allocation, Iris provides PyTorch-compatible functions for tensor creation and initialization. The library supports three categories of functions (see Table~\ref{lst:iris_host} for the complete list):

\begin{itemize}
    \item \textit{Creation functions}: \texttt{zeros}, \texttt{ones}, \texttt{empty}, \texttt{full}, \texttt{zeros\_like} for basic tensor initialization
    \item \textit{Range functions}: \texttt{arange} and \texttt{linspace} for generating sequences
    \item \textit{Random functions}: \texttt{rand}, \texttt{randn}, \texttt{randint}, \texttt{uniform} for sampling from various distributions
\end{itemize}

All functions support standard PyTorch parameters (\texttt{dtype}, \texttt{device}, \texttt{requires\_grad}), providing drop-in compatibility with existing PyTorch code. The key innovation is that all allocated tensors reside in the symmetric heap, enabling direct remote GPU access through Iris's device-side APIs.

\begin{table}[ht]
    \centering
    \caption{Iris Host-Side API Summary.}
    \label{lst:iris_host}
    \footnotesize
    \begin{tabular}{@{}p{0.15\columnwidth}p{0.25\columnwidth}p{0.52\columnwidth}@{}}
        \toprule
        \textbf{Category} & \textbf{Function}           & \textbf{Description}                                                \\
        \midrule
        \multirow{6}{*}{Core}
                          & \texttt{\_\_init\_\_()}     & Initialize Iris runtime, setup symmetric heap, exchange IPC handles \\
                          & \texttt{barrier()}          & Synchronize all ranks (GPU sync + distributed barrier)              \\
                          & \texttt{broadcast()}        & Broadcast tensor or scalar from one rank to all others              \\
                          & \texttt{get\_rank()}        & Return current process rank ID                                      \\
                          & \texttt{get\_num\_ranks()}  & Return total number of ranks (world size)                           \\
                          & \texttt{get\_heap\_bases()} & Return tensor containing all symmetric heap base addresses          \\
        \midrule
        \multirow{9}{*}{\parbox{2cm}{Tensor                                                                                   \\Creation}}
                          & \texttt{zeros()}            & Create tensor filled with zeros in symmetric heap                   \\
                          & \texttt{ones()}             & Create tensor filled with ones in symmetric heap                    \\
                          & \texttt{empty()}            & Create uninitialized tensor in symmetric heap                       \\
                          & \texttt{full()}             & Create tensor filled with specified value                           \\
                          & \texttt{zeros\_like()}      & Create zeros tensor matching input tensor's shape                   \\
                          & \texttt{rand()}             & Create tensor with uniform random values [0, 1)                     \\
                          & \texttt{randn()}            & Create tensor with normal distribution (mean=0, std=1)              \\
                          & \texttt{randint()}          & Create tensor with random integers in range                         \\
                          & \texttt{uniform()}          & Create tensor with uniform random values in [low, high)             \\
        \midrule
        \multirow{2}{*}{Sequences}
                          & \texttt{arange()}           & Create 1D tensor with evenly spaced values                          \\
                          & \texttt{linspace()}         & Create 1D tensor with linearly spaced values                        \\
        \bottomrule
    \end{tabular}
\end{table}

\begin{table}[ht]
    \centering
    \caption{Iris Device-Side API Summary}
    \label{lst:iris_device}
    \footnotesize
    \begin{tabular}{@{}p{0.15\columnwidth}p{0.25\columnwidth}p{0.52\columnwidth}@{}}
        \toprule
        \textbf{Category} & \textbf{Function}        & \textbf{Description}                                      \\
        \midrule
        \multirow{5}{*}{\parbox{2.5cm}{Memory                                                                    \\Operations}}
                          & \texttt{load()}          & Load value from remote rank's memory (value-based)        \\
                          & \texttt{store()}         & Store value to remote rank's memory (value-based)         \\
                          & \texttt{get()}           & Copy from remote memory to local memory (pointer-based)   \\
                          & \texttt{put()}           & Copy from local memory to remote memory (pointer-based)   \\
                          & \texttt{copy()}          & Copy between any two ranks (pointer-based)                \\
        \midrule
        \multirow{8}{*}{Atomics}
                          & \texttt{atomic\_add()}   & Atomically add value to remote memory location            \\
                          & \texttt{atomic\_xchg()}  & Atomically swap value with remote memory location         \\
                          & \texttt{atomic\_cas()}   & Atomically compare and conditionally swap if values match \\
                          & \texttt{atomic\_and()}   & Atomically perform bitwise AND on remote memory           \\
                          & \texttt{atomic\_or()}    & Atomically perform bitwise OR on remote memory            \\
                          & \texttt{atomic\_xor()}   & Atomically perform bitwise XOR on remote memory           \\
                          & \texttt{atomic\_min()}   & Atomically compute minimum with remote memory             \\
                          & \texttt{atomic\_max()}   & Atomically compute maximum with remote memory             \\
        \midrule
        Translation       & \texttt{\_\_translate()} & Internal function for pointer translation                 \\
        \bottomrule
    \end{tabular}
    \vspace{0.5em}
    \\
    \small
    \textit{Note:} All functions require \texttt{heap\_bases} parameter. Atomics support \texttt{sem} (relaxed/acquire/release/acq\_rel) and \texttt{scope} (block/gpu/sys).
\end{table}

\begin{listing}
    \noindent
    \begin{minted}[frame=lines, linenos, fontsize=\footnotesize,breaklines=true]{python}
@triton.jit
def load(pointer, to_rank, from_rank, heap_bases, mask=None):
    translated_ptr = __translate(pointer, to_rank, from_rank, heap_bases)
    result = tl.load(translated_ptr, mask=mask)
    return result

@triton.jit
def __translate(ptr, from_rank, to_rank, heap_bases):
    from_base = tl.load(heap_bases + from_rank)
    to_base = tl.load(heap_bases + to_rank)
    ptr_int = tl.cast(ptr, tl.uint64)
    offset = ptr_int - from_base
    to_base_byte = tl.cast(to_base, tl.pointer_type(tl.int8))
    translated_ptr_byte = to_base_byte + offset
    translated_ptr = tl.cast(translated_ptr_byte, ptr.dtype)
    return translated_ptr
\end{minted}
    \captionof{listing}{Iris load and pointer translation implementation.}
    \label{lst:iris_translate}
\end{listing}

\subsection{Device-Side APIs}

Iris provides Pythonic Triton-style device APIs for remote memory access and atomic operations. Since Triton doesn't support object-oriented programming, all functions require passing the symmetric heap pointer obtained via the \texttt{get\_heap\_bases} API\@. All device-side operations follow a consistent two-step pattern: (1) pointer translation from local to remote address space, and (2) memory operation on the translated pointer. Table~\ref{lst:iris_device} summarizes all device-side APIs.

\subsubsection{Pointer Translation Mechanism}

\begin{figure}[ht]
    \centering
    \includegraphics[width=0.8\columnwidth]{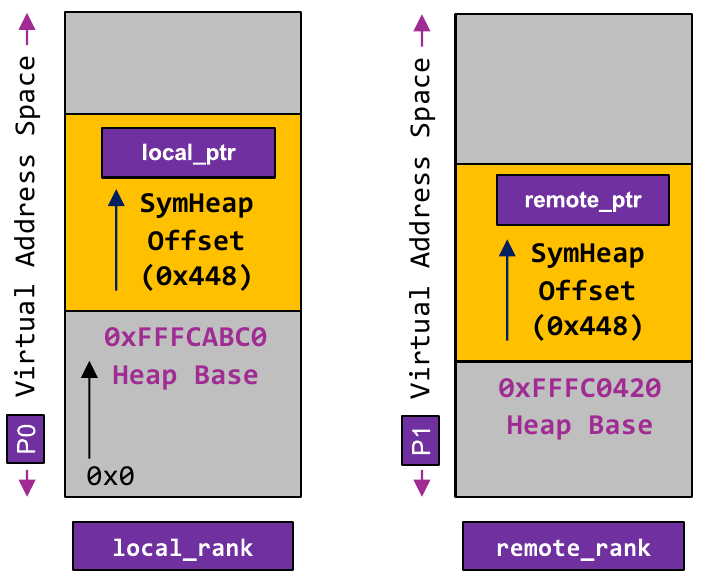}
    \caption{Pointer translation mechanism in Iris showing how local pointers are converted to remote addresses through offset computation.}
    \label{fig:translate_mechanism}
\end{figure}

The core of Iris's remote memory access is the \texttt{\_\_translate} function, which enables seamless access to remote memory without requiring explicit memory management from the programmer. The translation process (illustrated in Figure~\ref{fig:translate_mechanism}) works as follows:

\begin{enumerate}
    \item Compute the offset of the pointer within the local rank's symmetric heap
    \item Add this offset to the target rank's heap base address
    \item Cast the result back to the appropriate pointer type
\end{enumerate}

In the current version of Iris supporting intra-node communication via IPC, the remote operation can be directly performed on the computed remote pointer using standard memory operations (e.g., \texttt{tl.load}). While the implementation loads the \texttt{heap\_bases} on each call, experimental results show that these loads have no overhead, likely because the heap bases array (64 bytes) remains cached at the L1 level. Listing~\ref{lst:iris_translate} shows the implementation of \texttt{load} and \texttt{\_\_translate}, illustrating the two-step translation-then-operation pattern.

\subsubsection{Memory Operations}

Iris provides both value-based and pointer-based memory operations (see Table~\ref{lst:iris_device}):

\begin{itemize}
    \item \textit{Value-based operations}: \texttt{load()} and \texttt{store()} move data directly between registers and remote memory, enabling fine-grained register-to-memory transfers
    \item \textit{Pointer-based operations}: \texttt{get()}, \texttt{put()}, and \texttt{copy()} perform bulk transfers between memory regions, operating on buffer-to-buffer copies
\end{itemize}

All memory operations are non-blocking and use relaxed memory ordering by default. The choice between value-based and pointer-based operations depends on the communication pattern and data granularity requirements (discussed further in Section~\ref{sec:patterns}).

\subsubsection{Atomic Operations and Memory Model}

Iris provides the complete HIP/CUDA memory model semantics for atomic operations. The library supports three synchronization scopes:

\begin{itemize}
    \item \texttt{block}: synchronization visible within a thread block (CTA)
    \item \texttt{gpu}: synchronization visible across the entire GPU
    \item \texttt{sys}: synchronization visible system-wide across all GPUs
\end{itemize}

Memory ordering options include \texttt{relaxed}, \texttt{acquire}, \texttt{release}, and \texttt{acq\_rel}, enabling fine-grained control over synchronization semantics. The atomic operations include arithmetic (\texttt{add}), bitwise (\texttt{and}, \texttt{or}, \texttt{xor}), comparison (\texttt{min}, \texttt{max}), and exchange (\texttt{xchg}, \texttt{cas}) operations. All atomics follow the same pattern: translate the pointer, then perform the atomic operation with specified memory ordering and scope.

\subsubsection{Gluon Backend}
Iris also provides a Gluon backend that uses Triton's \texttt{@gluon.jit} decorator and \texttt{@aggregate} to encapsulate backend state, eliminating the need to pass \texttt{heap\_bases} manually. This backend offers improved ergonomics by encapsulating the Iris device context in an aggregate type. For example, instead of passing \texttt{heap\_bases} explicitly (standard API: \texttt{iris.load(buffer, to\_rank, from\_rank, heap\_bases)}), the Gluon API encapsulates this in a context object (\texttt{ctx.load(buffer, from\_rank=1)}). However, we focus this paper on the standard Triton API for clarity and broader compatibility.

\subsection{Iris Features and More}

\textbf{Tile-based APIs and programming model for communication.} Iris provides a tile-based communication model where operations are organized into tiles (e.g., \texttt{BLOCK\_SIZE\_M}, \texttt{BLOCK\_SIZE\_N}) that naturally fit within cache hierarchies. This tile-based approach enables building larger tiles within L1, L2, and LLC caches on chiplet-based architectures like MI300X and MI350X. Communication operations work at tile granularity, allowing fine-grained overlap where tiles can be communicated as soon as they are produced, rather than waiting for entire computation phases to complete. The tile-based model seamlessly integrates with Triton's blocked tensor operations, providing a unified programming model for both local computation and remote communication.

\textbf{Ease of instrumentation and profiling granularity within the kernel and library.} Since Iris is implemented entirely in Triton, profiling and instrumentation (e.g., through Triton's official profiler Proton) can be performed at fine-grained granularity both within user kernels and inside the Iris library itself. Developers can instrument specific communication operations, measure overlap efficiency, and analyze performance at the workgroup or even instruction level—not just at the boundaries of library calls, but deep within the library's implementation. This enables debugging and performance analysis of the pointer translation mechanism, remote memory operations, and synchronization primitives. This contrasts sharply with wrapped libraries where only opaque function call boundaries are visible, making it impossible to understand the performance characteristics of individual communication operations within a fused kernel or to diagnose issues inside the library itself.

\textbf{L1-, L2-, LLC-cache aware programming using swizzling and cache modifiers.} Iris provides explicit control over cache behavior through two complementary mechanisms. First, \textbf{cache modifiers} on load/store operations (e.g., \texttt{cache\_modifier=".wt"} for write-through) allow direct control over how data is written to memory hierarchy, enabling optimization for chiplet architectures where cache coherence across XCDs (Accelerator Complex Die) is critical. Second, \textbf{swizzling} is implemented at multiple levels: (1) across XCDs using \texttt{chiplet\_swizzle} to map work-groups to specific XCDs grouping tiles together for better Last-Level Cache (LLC) locality; and (2) spatial swizzling using \texttt{GROUP\_SIZE\_M} for L2-cache locality within tiles. This multi-level swizzling strategy, combined with cache modifiers, allows building larger tiles that efficiently utilize the entire cache hierarchy, from L1 caches within individual compute units to the LLC shared across XCDs.

\section{Building Complex Multi-GPU Patterns}
\label{sec:patterns}

Iris enables sophisticated distributed algorithms through a simple yet powerful API. To illustrate this, we present a taxonomy of fused and unfused compute-communication patterns, using General Matrix Multiplication (GEMM) and all-scatter as a case study.

GEMM is the foundational building block of modern GPU workloads, from deep learning to scientific simulation, responsible for most of the floating-point operations in large models. All-scatter, on the other hand, is a collective communication primitive where each GPU (or rank) distributes distinct portions of its data to all other GPUs. Together, they represent a common and challenging pattern: local compute producing partial results (via GEMM) that must be rapidly exchanged across devices (via all-scatter) to form the complete global output.

\begin{listing}
    \noindent
    \begin{minted}[frame=lines, linenos, fontsize=\footnotesize,breaklines=true]{python}
@triton.jit
def gemm_loop(A, B, C, ...):
    # Tile coordinate calculation removed for brevity
    # Memory layout setup
    rm = (pid_m * BLOCK_SIZE_M + tl.arange(0, BLOCK_SIZE_M)) % M
    rn = (pid_n * BLOCK_SIZE_N + tl.arange(0, BLOCK_SIZE_N)) % N
    rk = tl.arange(0, BLOCK_SIZE_K)
    A_BASE = A + rm[:, None] * stride_am + rk[None, :] * stride_ak
    B_BASE = B + rk[:, None] * stride_bk + rn[None, :] * stride_bn

    # Initialize accumulator registers
    acc = tl.zeros((BLOCK_SIZE_M, BLOCK_SIZE_N), dtype=tl.float32)

    # GEMM's Main loop
    for k in range( tl.cdiv(K, BLOCK_SIZE_K)):
        a = tl.load(tl.multiple_of(A_BASE, (1, 16)))
        b = tl.load(tl.multiple_of(B_BASE, (16, 1)))
        acc += tl.dot(a, b)
        A_BASE += BLOCK_SIZE_K * stride_ak
        B_BASE += BLOCK_SIZE_K * stride_bk

    # Non-even K handling removed for brevity
    ...

    # Accumulator registers with C results
    return acc.to(C.type.element_ty)
\end{minted}
    \captionof{listing}{A Triton GEMM main-loop routine, repurposed for several algorithms explained in the paper.}
    \label{lst:triton_gemm}
\end{listing}

With Iris, these patterns can be expressed naturally within Triton, developers can write kernels that overlap GEMM computation with communication, eliminating execution ``bubbles''\footnote{Bubbles refer to idle pipeline stages where no useful work occurs, often due to kernel launch overhead or synchronization delays.}. This overlap is notoriously difficult to achieve in practice, yet Iris makes it straightforward through intuitive APIs that integrate remote memory operations directly into the Triton programming model.

\begin{figure}
    \centering
    \includegraphics[width=\linewidth]{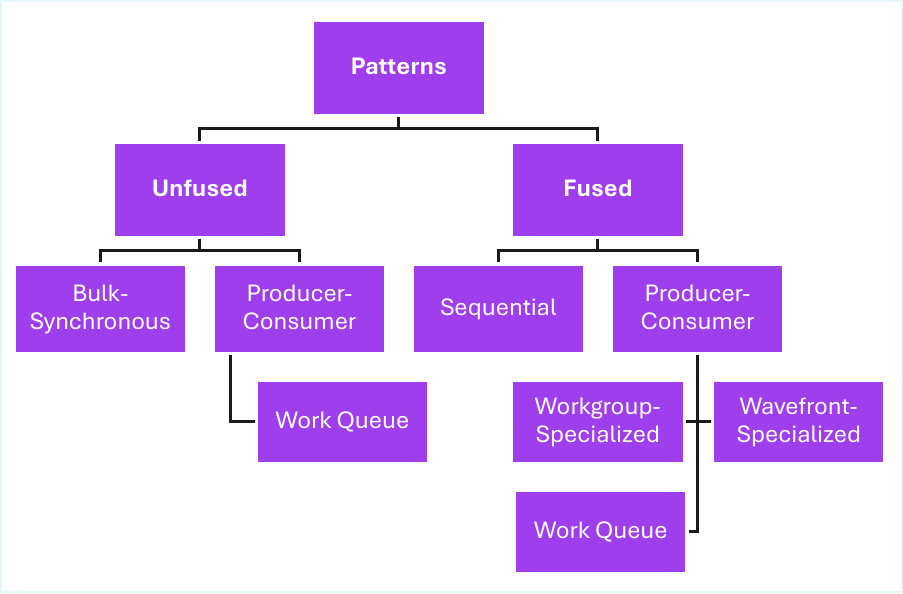}
    \caption{Taxonomy of unfused and fused computation and communication overlap patterns.}
    \label{fig:taxonomy}
\end{figure}

In contrast to the traditional complexity of building fused kernels, Iris enables developers to construct multi-GPU pipelines that are both efficient and maintainable. As we show next, this provides a practical taxonomy of compute-communication overlap strategies captured in Figure~\ref{fig:taxonomy}.

\begin{figure*}
    \centering
    \centering
    \includegraphics[width=\linewidth]{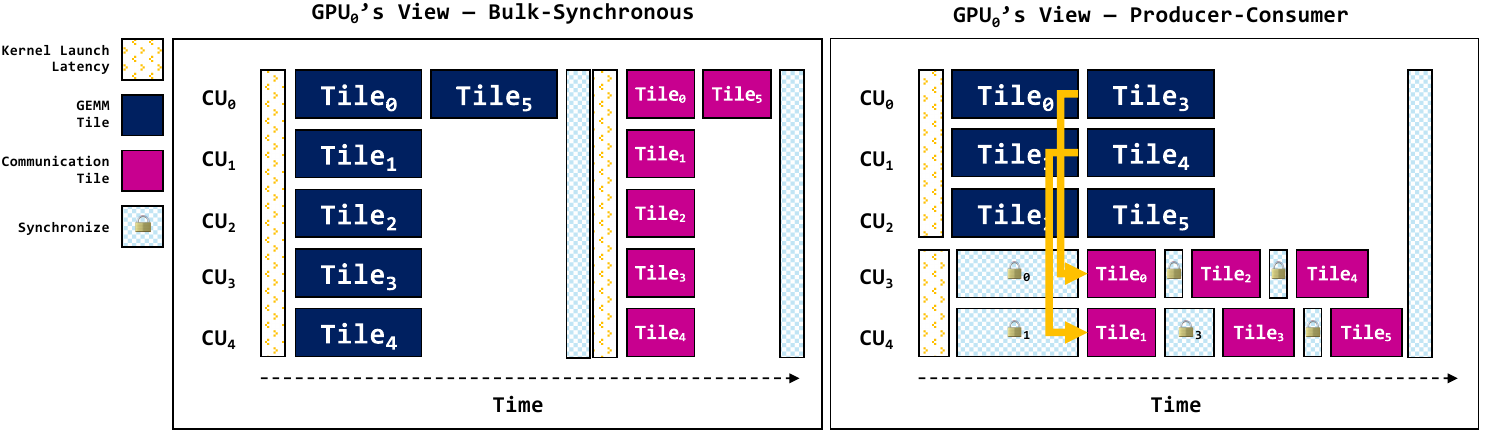}
    \caption{Timeline: Illustrates a single GPU's view of the taxonomy of unfused computation and communication patterns. (left) bulk-synchronous highlights the hard synchronization barriers that exist after each kernel, and (right) a multi-kernel producer-consumer pattern shows how overlap can be achieved by moving the synchronization at a finer granularity and partitioning the compute units (CUs) between computation and communication workers.}
    \label{fig:patterns_unfused}
\end{figure*}

\begin{figure*}
    \centering
    \includegraphics[width=\linewidth]{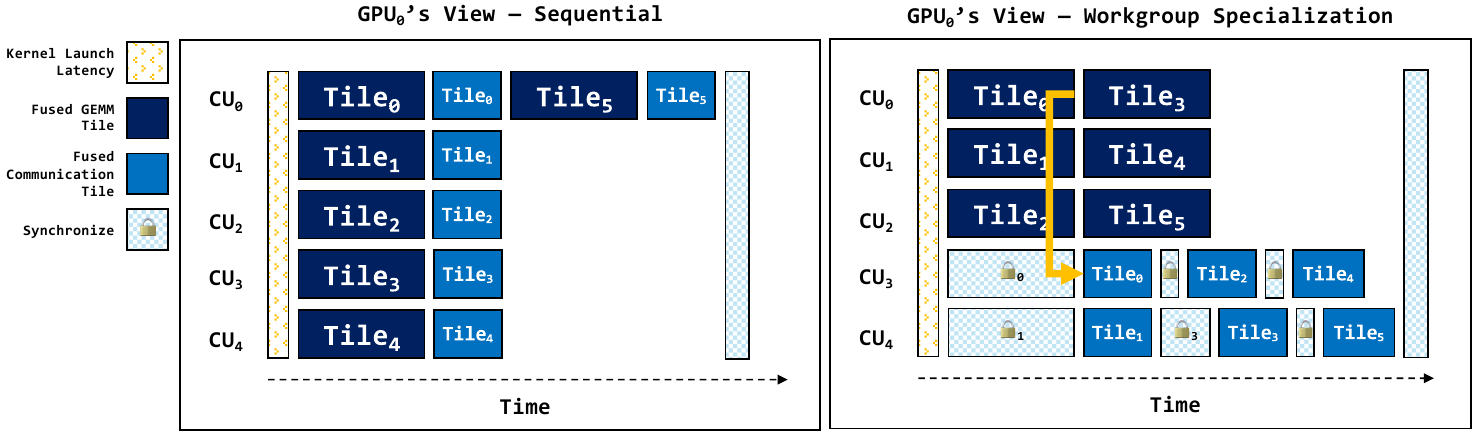}
    \caption{Timeline: Illustrates a single GPU's view of the taxonomy of fused GEMM and All-Scatter patterns.}
    \label{fig:patterns_fused}
\end{figure*}

We organize the patterns into two main categories: unfused patterns where computation and communication execute in separate kernels, and fused patterns where both operations are combined within a single kernel. Each category offers different trade-offs between implementation complexity, resource utilization, and performance characteristics.

\subsection{Unfused Patterns}

Unfused patterns separate computation and communication into distinct kernels, providing clear boundaries between operations. We begin with the simplest approach and progress to more sophisticated producer-consumer strategies.

\subsubsection{Bulk-Synchronous}

The simplest approach to coordinating compute and communication is the bulk-synchronous pattern, where operations execute sequentially with explicit synchronization barriers between kernels. In this pattern, the GEMM kernel first completes all computation and stores results to local GPU memory, and only after the entire GEMM kernel finishes and synchronizes does the all-scatter kernel begin, reading from local memory and distributing data to remote GPUs using \texttt{iris.put}. Listing~\ref{lst:iris_unfused_bsp} demonstrates this pattern: two separate kernels are launched sequentially on the same stream, establishing a strict data dependency that ensures the GEMM kernel fully completes before any communication begins.

This approach offers the benefit of simplicity and clear separation of concerns---each kernel has a single, well-defined responsibility. However, it introduces significant execution ``bubbles'' as shown in Figure~\ref{fig:patterns_unfused}: the GPU must wait for all GEMM work to complete and all workgroups to synchronize before any communication can proceed, leaving computational resources idle during the synchronization barrier. The pattern also requires intermediate writes to global memory, as the GEMM results must be stored before the all-scatter kernel can read them, adding memory bandwidth overhead that could be avoided with more sophisticated overlap strategies.

\begin{listing}
    \noindent
    \begin{minted}[frame=lines, linenos, fontsize=\footnotesize,breaklines=true]{python}
@triton.jit()
def gemm(
    A, B, C, ...
):
    pid = tl.program_id(0)
    for tile_id in range(pid, total_tiles, NUM_SMS):
        c = gemm_loop(A, B, C)
        ...
        # Store to local GPU's memory
        tl.store(C + offset, c, mask=mask, cache_modifier=".wt")

@triton.jit()
def all_scatter(C, ...):
    pid = tl.program_id(0)
    for tile_id in range(pid, total_tiles, NUM_SMS):

        # Begin: See the if segment for explanation:
        rm = (pid_m * BLOCK_SIZE_M + tl.arange(0, BLOCK_SIZE_M)) % M
        rn = (pid_n * BLOCK_SIZE_N + tl.arange(0, BLOCK_SIZE_N)) % N
        rm = tl.max_contiguous(tl.multiple_of(rm, BLOCK_SIZE_M), BLOCK_SIZE_M)
        rn = tl.max_contiguous(tl.multiple_of(rn, BLOCK_SIZE_N), BLOCK_SIZE_N)
        mask = (rm[:, None] < M) & (rn[None, :] < N)
        offset = rm[:, None] * stride_cm_global + (rn[None, :] + cur_rank * N) * stride_cn_global
        # End: masks/offset calculations.

        # Store from local to all other GPU's memory
        for remote_rank in range(world_size):
            if remote_rank != cur_rank:
                iris.put(C + offset, C + offset,
                cur_rank, remote_rank, heap_bases, mask=mask)

# On a single stream launch both kernels,
# establishing dependency of the two operations.
with torch.cuda.stream(main_stream):
    gemm[(num_sms,)](A, B, C, ...)
with torch.cuda.stream(main_stream):
    all_scatter[(num_sms,)](C, ...)
\end{minted}
    \captionof{listing}{Iris: Unfused, Bulk Synchronous -- illustrates the use of \texttt{iris.put} in a separate kernel after the GEMM kernel concludes and synchronizes.}
    \label{lst:iris_unfused_bsp}
\end{listing}

\subsubsection{Producer-Consumer (Stream Concurrency)}

Building on the bulk-synchronous pattern, the unfused producer-consumer approach achieves overlap by launching two separate kernels on asynchronous streams with explicit resource partitioning. Unlike bulk-synchronous execution where each kernel uses the entire GPU sequentially, this pattern limits the number of compute units (CUs) or streaming multiprocessors (SMs) allocated to each kernel. One kernel (the producer) uses a subset of CUs to perform GEMM computation, while another kernel (the consumer) uses the remaining CUs to perform communication. Dependencies between the kernels are managed through atomic-based synchronization primitives (similar to Listing~\ref{lst:workgroup_specialized}), but instead of using an if/else statement within a single fused kernel, the two operations execute as separate kernels on different streams. This approach enables concurrent execution of computation and communication while maintaining explicit control over resource allocation.

For example, on an MI300X with 304 compute units, the producer kernel might be launched with 256 SMs to handle GEMM tiles, while the consumer kernel runs concurrently with the remaining 48 SMs to perform all-scatter communication. The producer kernel writes completed tiles to local memory and signals their availability using atomic operations (e.g., \texttt{tl.atomic\_cas} with release semantics). The consumer kernel, executing concurrently on a separate stream, spins on these atomic locks (with acquire semantics) and immediately begins scattering tiles to remote GPUs as they become available. This pattern offers the modularity benefits of separate kernels while achieving overlap through hardware concurrency, though it requires careful tuning of the CU partition to balance computation and communication workloads.

\subsection{Fused Patterns}

While unfused patterns provide simplicity and modularity, fused patterns offer superior performance by eliminating kernel launch overhead and enabling fine-grained computation-communication overlap within a single kernel. These patterns leverage Iris's native Triton implementation to seamlessly interleave operations at tile granularity.

\subsubsection{Sequential}

\begin{listing}
    \noindent
    \begin{minted}[frame=lines, linenos, fontsize=\footnotesize,breaklines=true]{python}
@triton.jit
def fused_gemm_all_scatter(
    A, B, C, ...
):
    pid = tl.program_id(0)
    num_pid_m, num_pid_n = tl.cdiv(M, BLOCK_SIZE_M), tl.cdiv(N, BLOCK_SIZE_N)
    total_tiles = num_pid_m * num_pid_n
    for tile_id in range(pid, total_tiles, NUM_SMS):
        c = gemm_loop(A, B, C)

        rm = (pid_m * BLOCK_SIZE_M + tl.arange(0, BLOCK_SIZE_M)) % M
        rn = (pid_n * BLOCK_SIZE_N + tl.arange(0, BLOCK_SIZE_N)) % N

        # Add compiler hints
        rm = tl.max_contiguous(tl.multiple_of(rm, BLOCK_SIZE_M), BLOCK_SIZE_M)
        rn = tl.max_contiguous(tl.multiple_of(rn, BLOCK_SIZE_N), BLOCK_SIZE_N)

        # Define the C-mask (BLOCK_SIZE_M, 1) x (1, BLOCK_SIZE_N)
        mask = (rm[:, None] < M) & (rn[None, :] < N)

        # Calculate the "global" offset of C based on the rank.
        # Note the N-dimension is being multiplied by current rank.
        # This is because each rank is computing a portion of the
        # N-dimension locally and then scattering it to all other
        # ranks to complete the global N-dimension.
        offset = rm[:, None] * stride_cm_global + (rn[None, :] + cur_rank * N) * stride_cn_global

        # Scatter to all ranks
        for remote_rank in range(world_size):
            iris.store(C + offset, c, cur_rank, remote_rank, heap_bases, mask=mask)
\end{minted}
    \captionof{listing}{Iris: Fused, Sequential -- illustrates the use of \texttt{iris.store} right after the GEMM tile is produced.}
    \label{lst:iris_allscatter}
\end{listing}

To bring more control of scheduling the work in the hands of developers, we can fuse multiple operations together in a mega or uber kernel~\cite{DeepSeekAI:2025:DVT,Spector:2025:LMN} and reduce the overhead of tearing down and recreating the kernels. This approach significantly reduces the ``bubbles'' in the total workload by moving to a fine-grained synchronization approach at a tile granularity. One such way fused kernels are implemented is following the data-dependencies that inherently exist within those operators, for example, we insert the all-scatter operator sequentially after the computation of each output GEMM tile. Iris enables this pattern through its device-side APIs that operate directly within Triton kernels. Listing~\ref{lst:iris_allscatter} illustrates how developers may use \texttt{iris.store} to immediately scatter the GEMM tile produced to all remote GPUs without the need for a bulk-synchronous kernel-level barrier (see Listing~\ref{lst:iris_unfused_bsp}).

Such pattern has the benefit of operating on the data as soon as it is ready (such as all scatter's store on the accumulator registers) with no intermediate writes to global memory required. Fused operators, however, still retain the sequential dependencies of executing one operator, waiting for example, the GEMM to complete and issuing the next operator in the same kernel. The impact of tail latency (tail occupancy inefficiency) worsens, because now the last ``wave'' of work needs to process GEMM and all scatter before the kernel completes.

\subsubsection{Producer-Consumer (Workgroup Specialization)}

One such way of avoiding the sequential issuance of the two operator is by using specialization techniques over the available compute resources. We can implement a persistent-style kernels (see the for-loop over all tiles in Listings~\ref{lst:iris_allscatter}) and specialize the type of computation each compute resource (e.g., workgroups) does by using the workgroup index (\texttt{pid} in Triton). Listing~\ref{lst:workgroup_specialized} shows an example of GEMM + All Scatter using Iris, where each workgroup gets mapped to a compute unit of an AMD GPU (MI300X has 304 compute units); the first 256 (0-255) workgroups are responsible for computing the GEMM output tile and signaling the other 48 workgroups (256-303) responsible for waiting for a tile to be produced (using for example a spin-lock), and then scatter the result to other GPUs. With this method, we can dedicate exact compute resources for various tasks---this is especially useful when workloads like GEMMs do not require the entire device to achieve peak performance. Fused workgroup specialization, however, just like all other fused kernels, requires worst-case resource allocation (i.e., an operation such as all-scatter is forced to occupy more resources than needed because it is fused with a more resource-intensive operation such as GEMM).

\begin{listing}
    \noindent
    \begin{minted}[frame=lines, linenos, fontsize=\footnotesize,breaklines=true]{python}
@triton.jit()
def wg_specialized_gemm_all_scatter(
    A, B, C, locks, GEMM_SMS, COMM_SMS, ...
):
    pid = tl.program_id(0)
    if pid < GEMM_SMS:

        # Process all gemm tiles using GEMM_SMS number of
        # workgroups in a persistent fashion.
        for tile_id in range(pid, total_tiles, GEMM_SMS):
            c = gemm_loop(A, B, C)
            ...
            # Store to local GPU's memory
            tl.store(C + offset, c, mask=mask, cache_modifier=".wt")
            tl.atomic_cas(locks + tile_id, 0, 1, sem="release", scope="gpu")

    else:  # pid >= GEMM_SMS
        COMM_SMS = NUM_SMS - GEMM_SMS
        pid = pid - GEMM_SMS

        # Process all comm tiles using COMM_SMS number of
        # workgroups in a persistent fashion.
        for tile_id in range(pid, total_tiles, COMM_SMS):

            # Wait for the tile to be ready.
            while tl.atomic_cas(locks + tile_id, 1, 0, sem="acquire", scope="gpu") == 0:
	           pass

            # Store from local to all other GPU's memory
            for remote_rank in range(world_size):
                if remote_rank != cur_rank:
                    iris.put(C + offset, C + offset,
                    cur_rank, remote_rank, heap_bases, mask=mask)

# Launch code:
with torch.cuda.stream(main_stream):
    wg_specialized_gemm_all_scatter[(num_sms,)](
        A, B, C, locks, GEMM_SMS, COMM_SMS, ...)
\end{minted}
    \captionof{listing}{Iris: Fused, Workgroup Specialization -- illustrates how a single fused kernel can be split into components where dedicated workgroups perform either communication or computation operations.}
    \label{lst:workgroup_specialized}
\end{listing}

\subsubsection{Producer-Consumer (Wave Specialization) and Work Queue}

Similar to workgroup specialization, we can also split the work at a finer granularity of a wavefront (AMD GPU) or a warp (NVIDIA GPU), where 64 or 32 threads in a lockstep fashion work on issuing communication or processing compute. However, without using Gluon, this pattern is not typically suited for a more workgroup-centric language like Triton. Work queue on the other hand extends these patterns and moves the management of the work in a separate queue-like data structure, due to the synchronization cost of inserting and removing ``work'' (communication or computation tile), queues are also not well suited for a GPU architecture (or Triton language.) In this paper, we focus on all other patterns described in the previous subsections.
\section{Results}
\label{sec:results}

We evaluate Iris on a system with 8x AMD Instinct\texttrademark~MI300X GPUs configured under NPS1/SPX memory and compute partition modes~\cite{Osama:2025:DDM} and ROCm 6.3.1. Our evaluation consists of two main components: microbenchmarks that characterize Iris's fundamental performance across point-to-point and collective communication primitives, and real-world application studies using the GEMM+All-Scatter fused patterns from Section~\ref{sec:patterns}. The microbenchmarks demonstrate that Iris achieves near-optimal bandwidth utilization across all operations, validating the efficiency of its native Triton implementation. For real-world workloads, Iris's fine-grained overlap capabilities enable significant performance improvements over the bulk-synchronous baseline, with speedups ranging from 0.93$\times$ to 1.79$\times$ (average 1.21$\times$) compared to PyTorch and RCCL. These results highlight both the low overhead of Iris's abstraction and the substantial benefits of tile-granularity computation-communication overlap enabled by its design.

\subsection{Microbenchmarks}

\begin{figure*}
    \centering

    \begin{subfigure}[t]{0.48\textwidth}
        \centering
        \includegraphics[width=\linewidth]{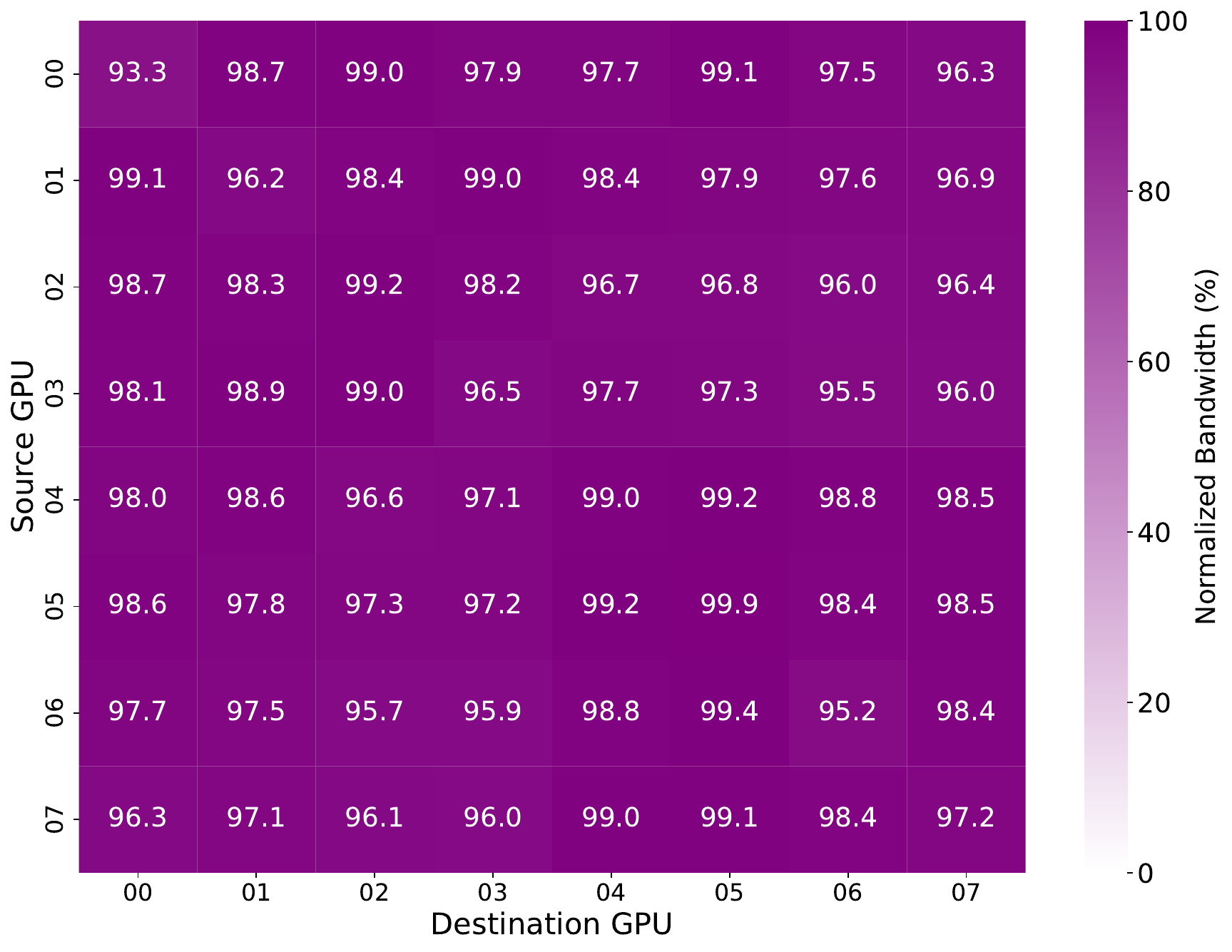}
        \caption{Load benchmark}
        \label{fig:load_bench}
    \end{subfigure}
    \hfill
    \begin{subfigure}[t]{0.48\textwidth}
        \centering
        \includegraphics[width=\linewidth]{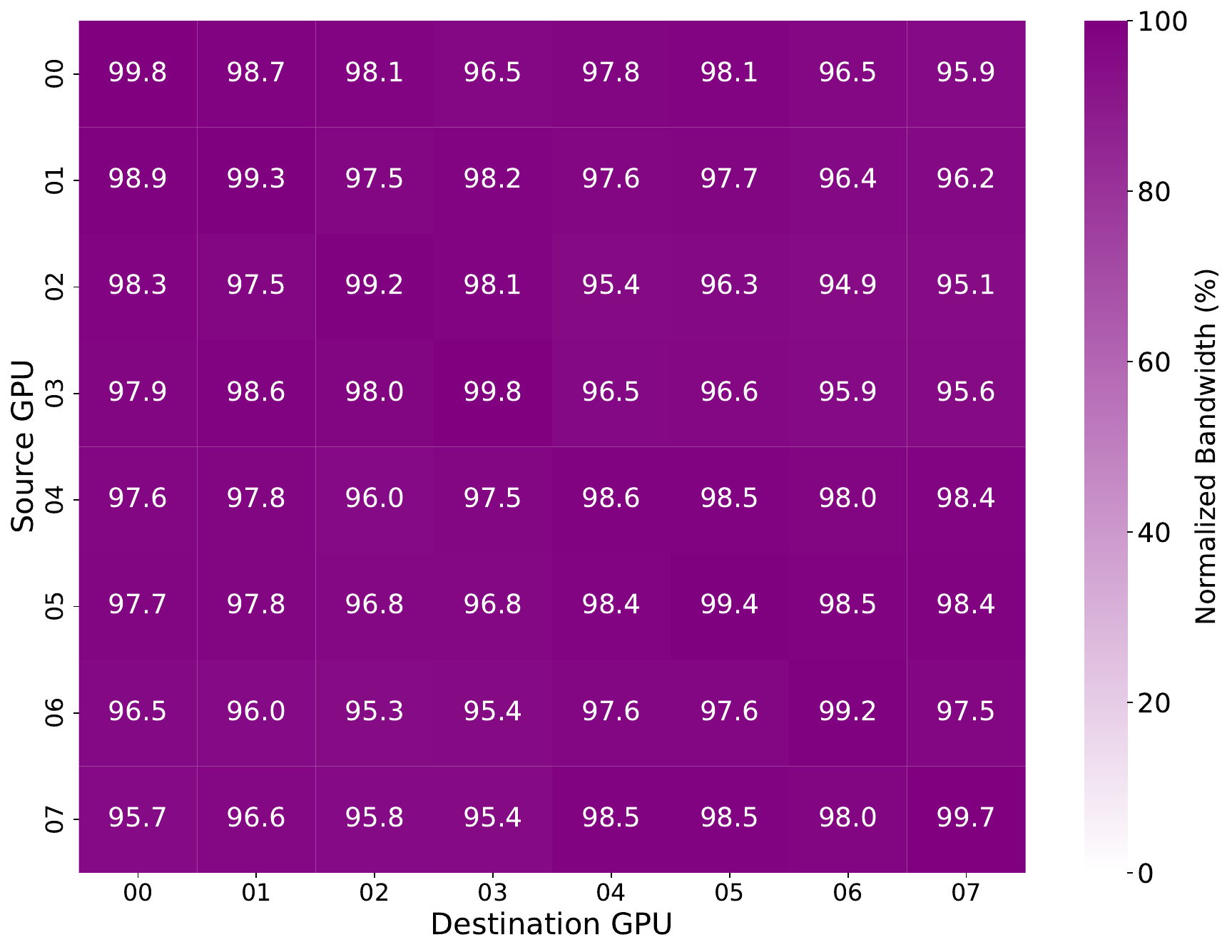}
        \caption{Store benchmark}
        \label{fig:store_bench}
    \end{subfigure}

    \vspace{0.5em}

    \begin{subfigure}[t]{0.48\textwidth}
        \centering
        \includegraphics[width=\linewidth]{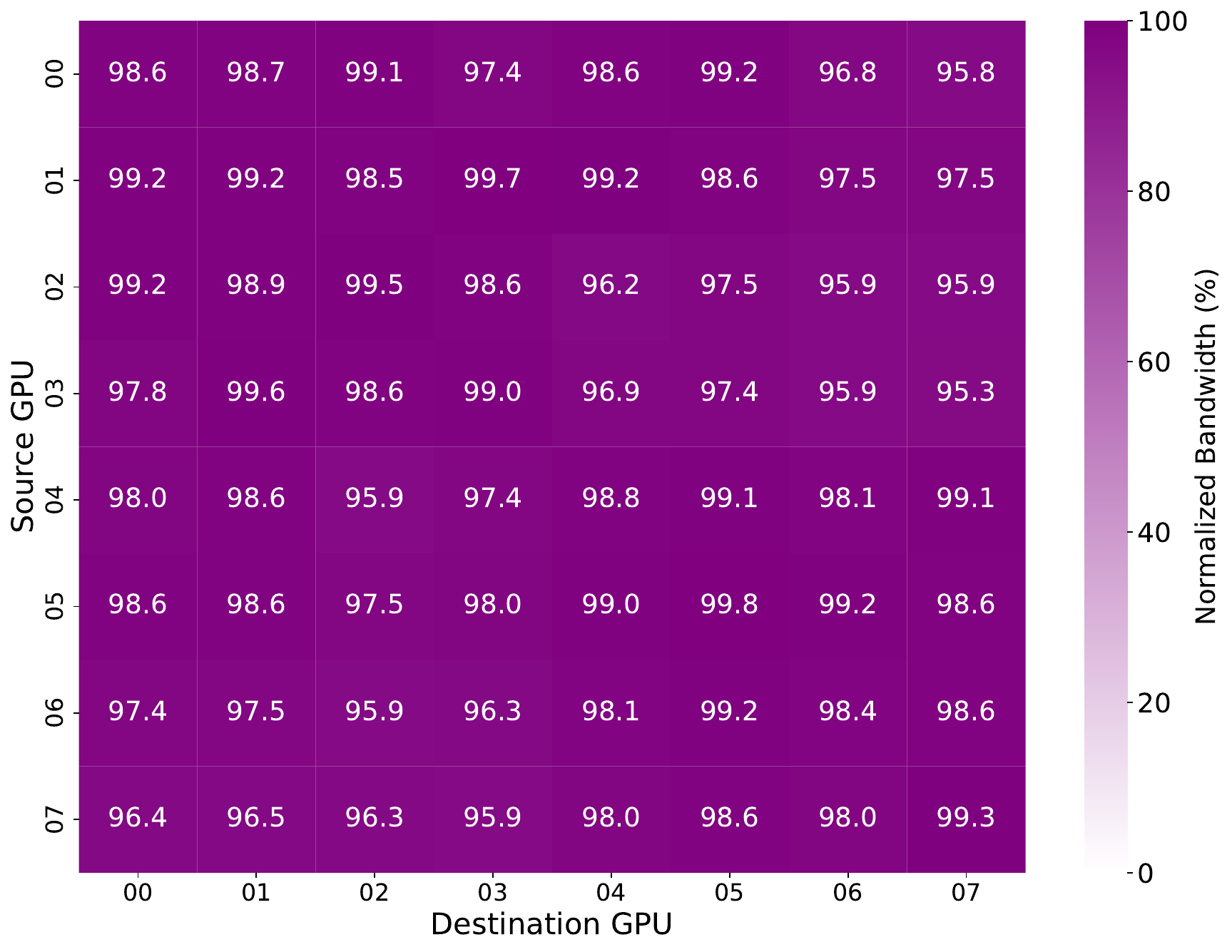}
        \caption{Atomic add}
        \label{fig:atomic_add}
    \end{subfigure}
    \hfill
    \begin{subfigure}[t]{0.48\textwidth}
        \centering
        \includegraphics[width=\linewidth]{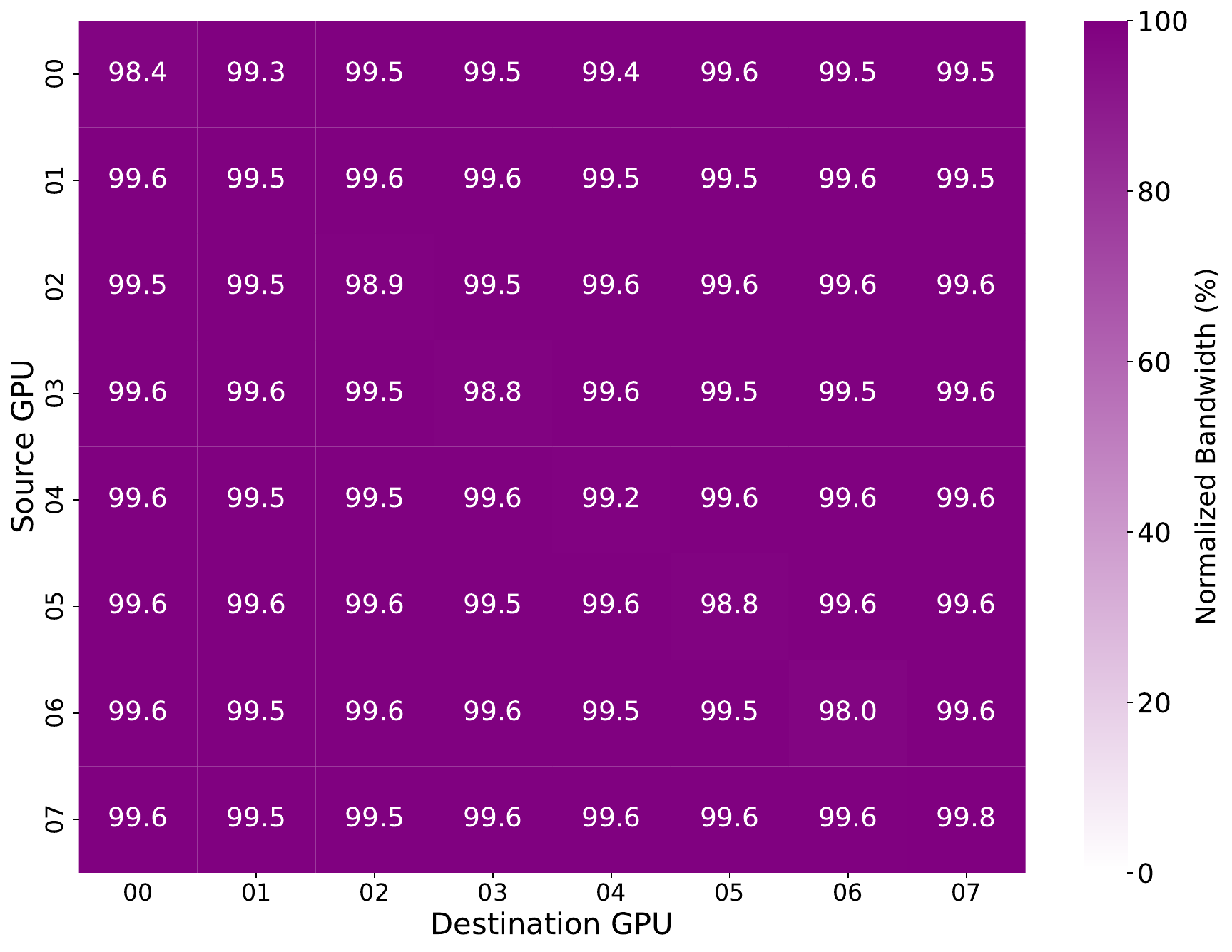}
        \caption{Atomic xchg}
        \label{fig:atomic_xchg}
    \end{subfigure}

    \caption{Performance benchmarks for load, store, and atomic operations. The results demonstrate Iris's efficiency in handling different types of remote memory access patterns.}
    \label{fig:load_store_atomics}
\end{figure*}

Figure~\ref{fig:load_store_atomics} presents performance benchmarks for point-to-point load, store, and atomic operations. All benchmarks are normalized relative to the achievable bandwidth~\cite{Sander:2025:UPM,Sander:2025:MMA} on the system, with the heatmaps showing bandwidth percentages where darker shades indicate better (higher) performance. The results demonstrate Iris's efficiency in handling different types of remote memory access patterns, with consistent performance across operations, achieving near-optimal bandwidth utilization.

\begin{figure*}
    \centering
    \begin{subfigure}[b]{0.48\textwidth}
        \centering
        \includegraphics[width=\linewidth]{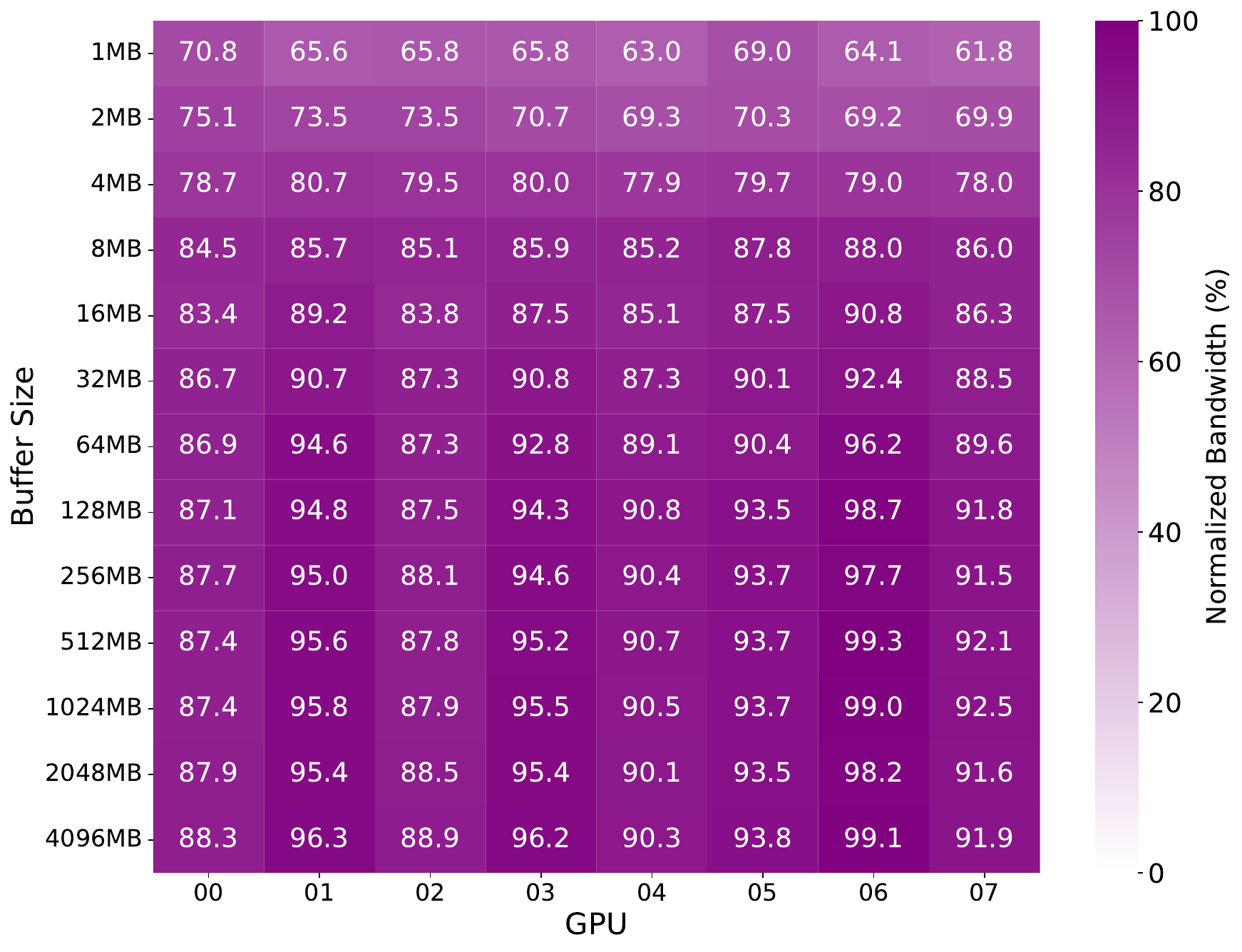}
        \caption{All-Load benchmark}
        \label{fig:all_load_bench}
    \end{subfigure}
    \hfill
    \begin{subfigure}[b]{0.48\textwidth}
        \centering
        \includegraphics[width=\linewidth]{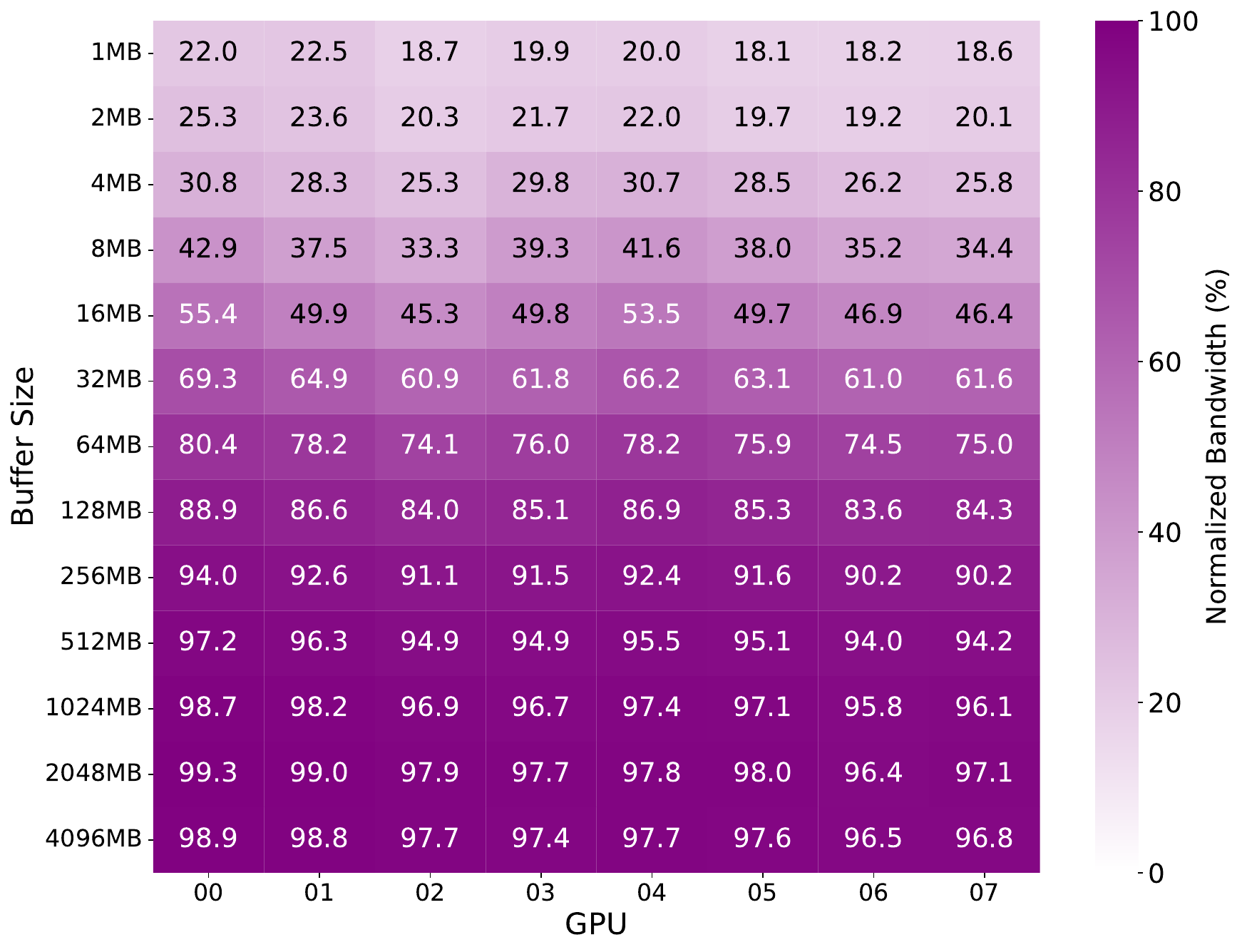}
        \caption{All-Store benchmark}
        \label{fig:all_store_bench}
    \end{subfigure}
    \caption{All load/all store benchmark results where all GPUs participate in all load/store operations across all links. As buffer size increases, performance approaches peak bandwidth utilization as expected.}
    \label{fig:all_load_store_comprehensive}
\end{figure*}

Figure~\ref{fig:all_load_store_comprehensive} provides the all-load and all-store mircobenchmark results where all GPUs participate in the load and store operations across all links. In the benchmark, different buffer sizes are moved across all ranks at the same time. The heatmaps show the normalized bandwidth relative to achievable bandwidth where darker shades represent superior performance. As buffer size increases, performance improves significantly (as expected), reaching near-optimal achievable bandwidth utilization showcasing the efficiency of the Iris simple-yet-effective implementation and abstraction.

\subsection{Evaluating Fused, Unfused Patterns Taxonomy}

\begin{figure*}
    \centering
    \includegraphics[width=\textwidth]{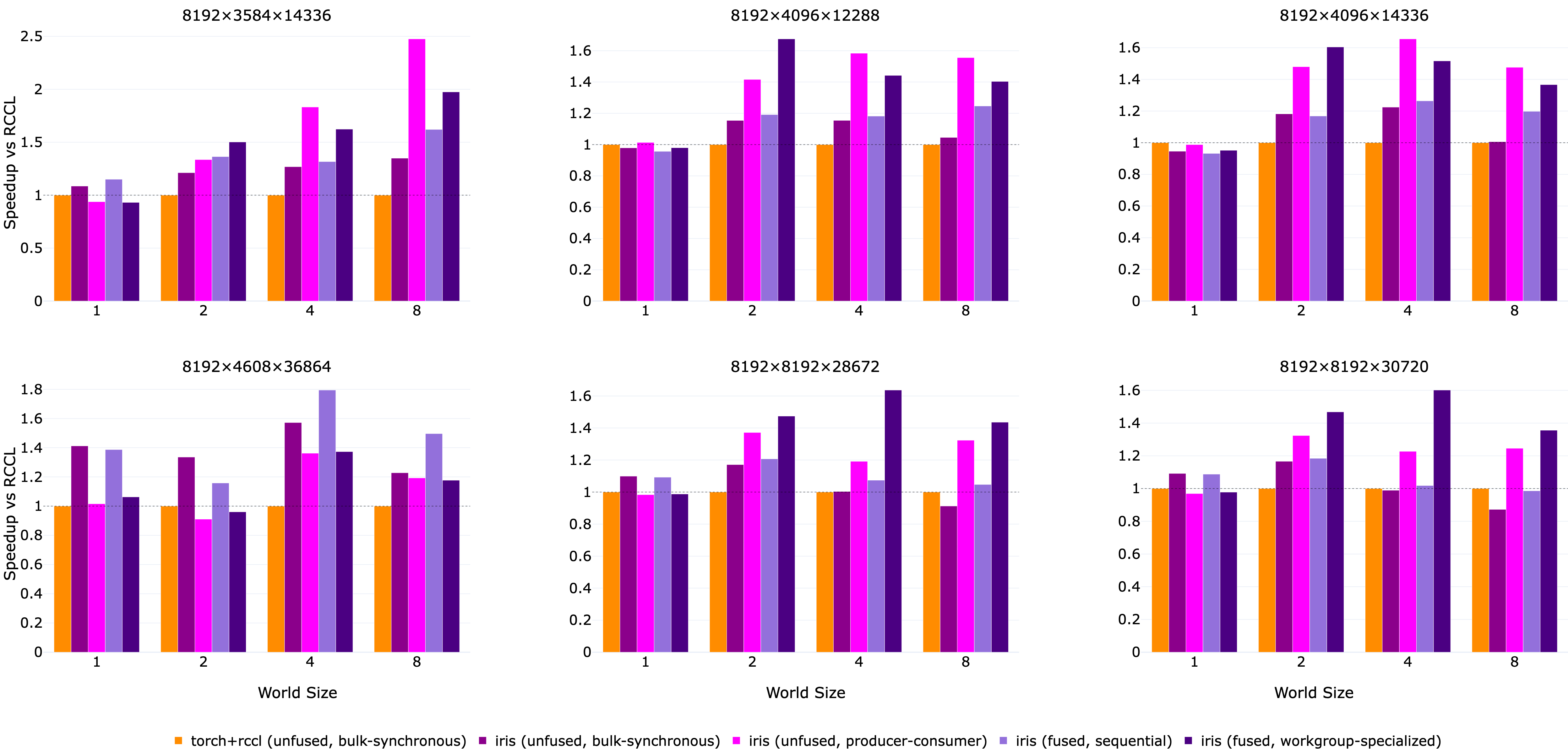}
    \caption{Complete performance comparison between Iris fused GEMM + All-Scatter and RCCL GEMM + AllGather kernels across different problem sizes and world sizes.}
    \label{fig:iris_vs_torch}
\end{figure*}

To evaluate Iris' in a real-world application, we continue our case-study from Section~\ref{sec:patterns}. We implemented many of the fused and unfused patterns described in the Figure~\ref{fig:taxonomy} and Listings~\ref{lst:iris_unfused_bsp},~\ref{lst:iris_allscatter} and~\ref{lst:workgroup_specialized} to capture the versatility of the abstraction and APIs. In this section, we cover a deep-dive of these patterns using PyTorch's \texttt{torch.matmul} and RCCL's All-Gather as a functionally equivalent baseline. Figure~\ref{fig:iris_vs_torch} shows the complete performance landscape across six different problem shapes and sizes with varying N and K-dimensions (and M=8192) for different number of GPUs (world size).

\begin{figure}
    \centering
    \includegraphics[width=\columnwidth]{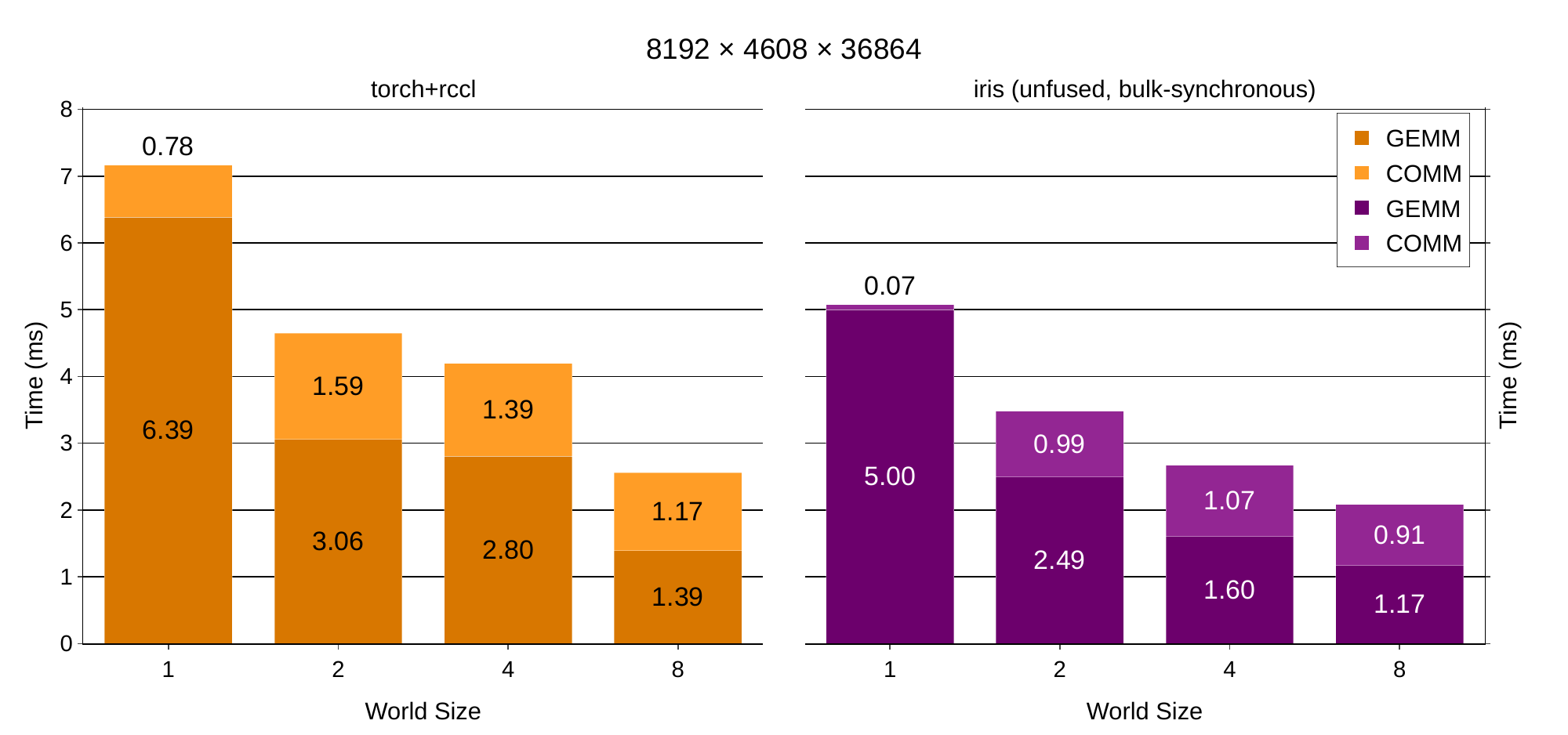}
    \caption{Deep-dive: Shows breakdown of GEMM (darker region) and Communication (lighter region) for Iris compared to PyTorch and RCCL for $8192 \times 4608 \times 36864$ matrix size. Note the slight speedups in both GEMM and communication due to Iris' flexibility to be able to cater to a specific problem shape and size. Tile-based abstraction allows for users to simply adjust the needed tile-size at compile-time per a problem/kernel granularity.}
    \label{fig:deepdive:bsp}
\end{figure}

We first establish Iris' baseline using the ``Unfused,\linebreak Bulk-synchronous'' schedule, which as a schedule is equivalent PyTorch and RCCL. We observe that Iris competes with state-of-the-art GEMM and All-Gather implementation --- this further validates that Iris' abstraction isn't resulting in any discernible overheads. In some cases, such as $8192 \times 4608 \times 36864$, Iris is 20\% faster using 8 GPUs. This is largely attributed to difference in heuristics for the PyTorch and RCCL's heuristics selecting a suboptimal configuration (see Figure~\ref{fig:deepdive:bsp}).

\begin{figure}
    \centering
    \includegraphics[width=\columnwidth]{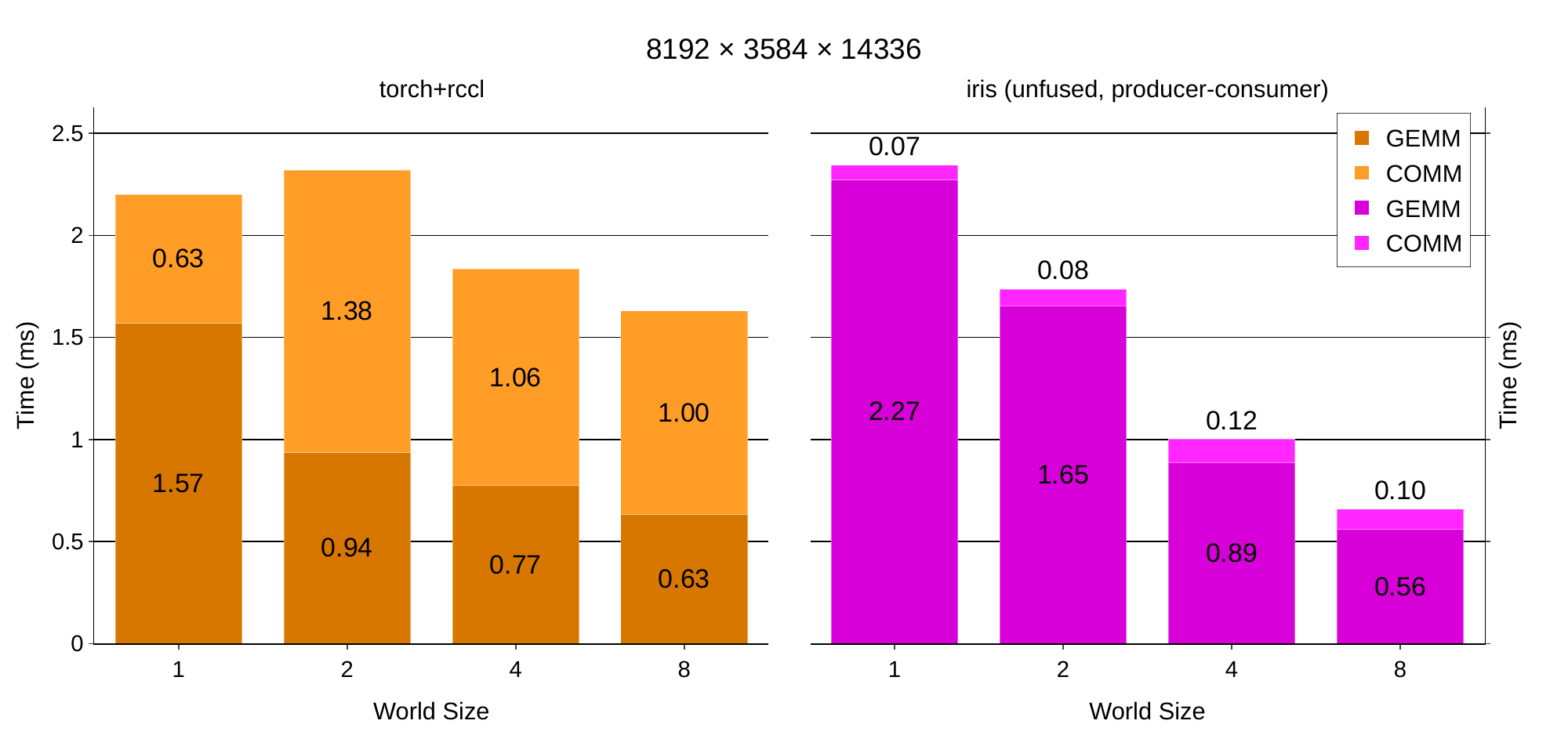}
    \includegraphics[width=\columnwidth]{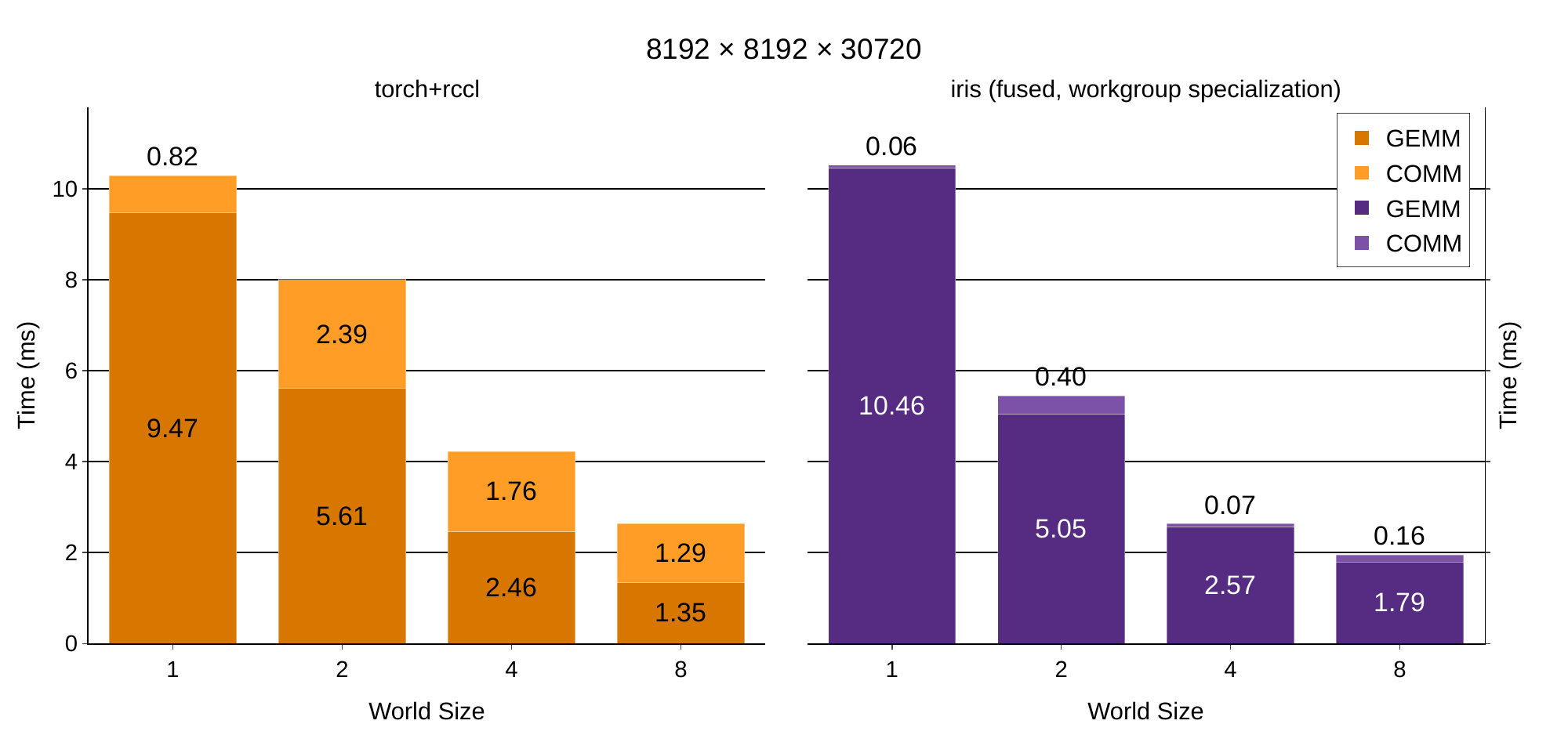}
    \caption{Deep-dive: Shows breakdown of GEMM (darker region) and Communication (lighter region) for Iris compared to PyTorch and RCCL for $8192 \times 3584 \times 14336$ matrix size. These two problems are specifically of interest for producer-consumer and workgroup specialization based schedules as they show the approach mostly hides the overhead of communication behind the GEMM (darker region) by splitting the available GPU's compute units between GEMM and communication.}
    \label{fig:deepdive:pc}
\end{figure}

Iris also allows to break the rigid bulk-synchronous programming model by using the device-side APIs and moving the synchronization at a fine grained tile-level granularity. ``Unfused, producer-consumer'', ``fused, workgroup-specialization'' and ``fused, sequential'' all follow this model. Unfused, producer-consumer approach gives up-to $2.5\times$ speedup for problem shape of $8192 \times 3584 \times 14336$ on 8~GPUs. The nature of the problem (small-N after being split 8-ways, and large-K) allows producer-consumer-style model (unfused or fused using workgroup-specialization) to completely hide the communication operation behind the GEMM operation, this is illustrated in Figure~\ref{fig:deepdive:pc}. The difference between the fused and unfused variants of producer-consumer approach is that using unfused two kernels (one producer and one consumer), we avoid worst-case resource allocation\footnote{Worst-case resource allocation: Size of the allocated shared-memory, number of VGPRs or number of threads launched is not bounded by the worst-case operation.} (typically bounded by GEMMs) and promotes better occupancy at the cost of kernel launch latency and less control over the scheduling of operations. Whereas in a fused variant, we only launch one kernel and do not have to pay an additional cost to launch the kernel, it allows for more scheduling control and re-purposing/reusing resources and data when relevant, however, requires worst-case resource allocation and limits the occupancy for one of the operator.

\begin{figure}
    \centering
    \includegraphics[width=\columnwidth]{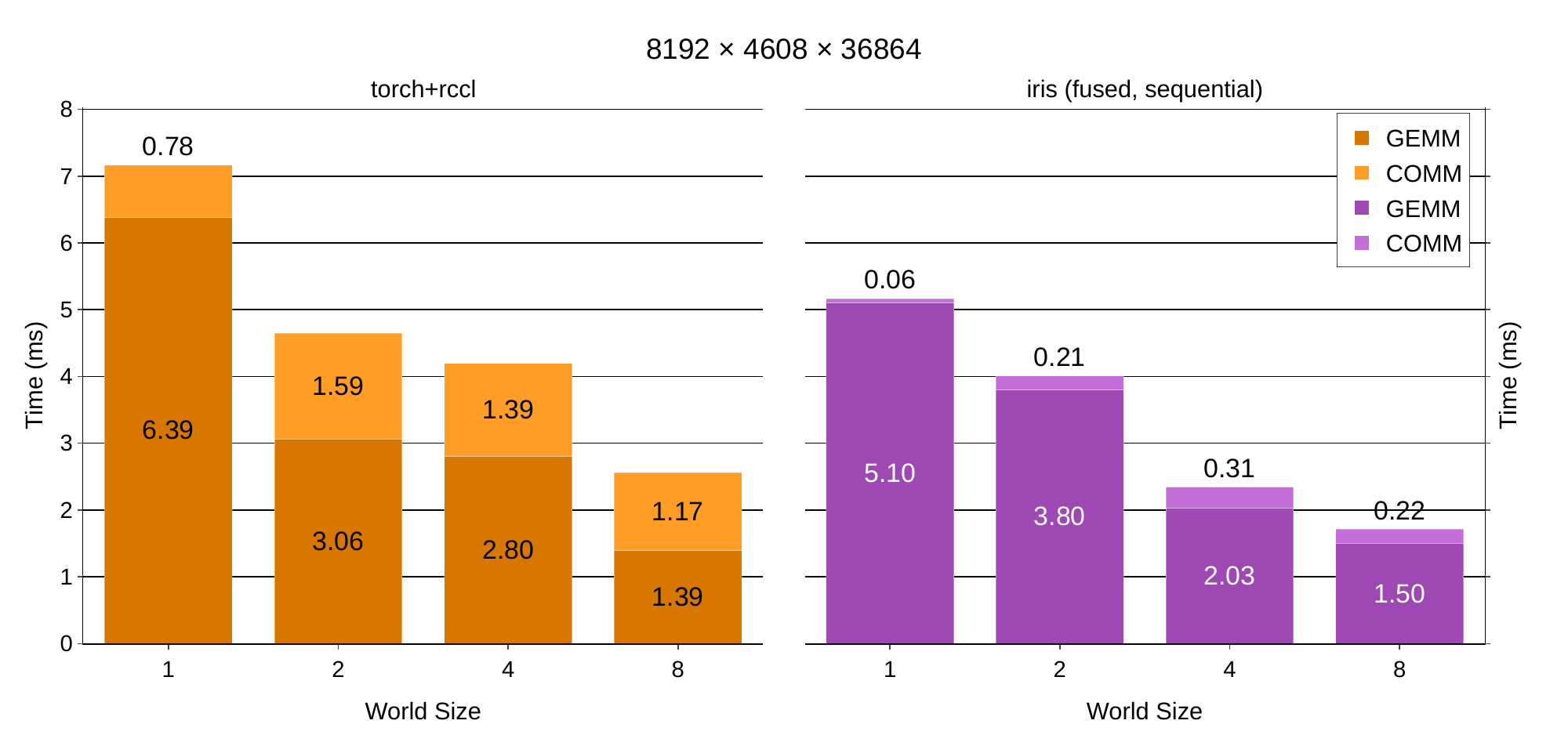}
    \caption{Deep-dive: Shows breakdown of GEMM (darker region) and Communication (lighter region) for Iris compared to PyTorch and RCCL for $8192 \times 4608 \times 36864$ matrix size. This shape illustrates when fused sequential approach benefits when the added communication is an overall small overhead (smaller output tile and really large K) and more resources can simply be allocated to processing GEMM.}
    \label{fig:deepdive:seq}
\end{figure}

A fused, sequential schedule is the simplest of them all --- essentially appending the communication operation at the end of the main-loop of the GEMM operation. This required a few lines of code changes as shown in Listing~\ref{lst:iris_allscatter}, and works well for problems that need more resources for GEMM (such as small N and massive large-K.) However, as the name suggest, this sort of schedule creates a sequential dependency between the GEMM operation and All-Scatter operation, and increases the tail latency of the entire problem. Potentially creating large bubbles in the last timestep of the problem (also illustrated in Figure~\ref{fig:patterns_fused}.) With this schedule, Iris outperforms the baseline PyTorch and RCCL implementation by $1.8\times$ for 4 GPUs and $1.5\times$ for 8 GPUs on $8192 \times 4608 \times 36864$ problem size (see Figure~\ref{fig:deepdive:seq}).

Across all tested configurations, Iris achieves an average speedup of 1.21$\times$ over PyTorch and RCCL, with speedups ranging from 0.93$\times$ to 1.79$\times$, highlighting the consistent performance benefits of Iris's design. These speedups are largely attributed to the flexibility of Iris' design to be able to implement a fused kernel with tile-granularity synchronization (and the resultant compute and communication overlap) versus a rigid, bulk-synchronous programming model of PyTorch's GEMM and RCCL's AllGather kernels.

\section{Conclusion and Future Work}

Iris's ability to match or exceed the performance of heavily-optimized HIP/CUDA-based libraries like RCCL despite its pure Triton implementation demonstrates that low-level native implementations are not a fundamental requirement for multi-GPU programming. The 1.79$\times$ peak speedup reflects a qualitatively different programming model where synchronization granularity shifts from kernel boundaries to tiles. Perhaps more significant is what Iris makes tractable: our taxonomy demonstrates that diverse overlap patterns—from bulk-synchronous to fused sequential to producer-consumer to workgroup specialization—can be implemented with minimal code changes, often requiring only a few additional lines within the same Triton kernel. Patterns that would demand substantial engineering effort with traditional CCLs (separate kernel implementations, complex host-side coordination, manual resource partitioning) emerge naturally in Iris using the same primitives Triton developers already use for single-GPU work.

This suggests that the real barrier to fine-grained overlap has been abstraction mismatch, not hardware capability. When communication primitives live in the same semantic space as computation (tile-based Triton), overlap patterns become straightforward extensions rather than heroic engineering efforts. Future work will focus on extending Iris to multi-node settings with RDMA, exploring additional fused patterns such as wave-specialization in Gluon and work queues, and investigating opportunities to offload Iris operations to the compiler itself, leveraging Triton's ability to optimize across the entire computation-communication pipeline.

\begin{acks}
    This work was supported in part by Advanced Micro Devices, Inc. under the AMD AI \& HPC Cluster Program. The authors would like to thank Karl Schulz, Lei Zhang, Ziyad AlBanoby, Octavian-Alexandru Trifan, David Sidler, Xiaohu Guo, Karthik Sangaiah, Lixun Zhang, Vinayak Gokhale, Panagiotis Mylonas, Eric Eaton, Aditya Nandakumar, Ahmed Eltantawy, Dimple Prajapati, Mike Chu, Mike Schulte, Ganesh Dasika, Brad Beckmann, Ralph Wittig, and Peng Sun for their continuous feedback, support and suggestions. AMD, the AMD Arrow logo, AMD CDNA\texttrademark, AMD Instinct\texttrademark, AMD ROCm\texttrademark,
    AMD Infinity Cache\texttrademark, AMD Infinity Fabric\texttrademark, and combinations
    thereof are trademarks of Advanced Micro Devices, Inc\@. Other
    product names used in this publication are for identification
    purposes only and may be trademarks of their respective
    companies.

\end{acks}

\bibliographystyle{ACM-Reference-Format}
\bibliography{iris}

\end{document}